\documentclass[american,english,aps,manuscript]{revtex4}
\usepackage[T1]{fontenc}
\usepackage[latin9]{inputenc}
\usepackage{float}
\usepackage{amssymb}
\usepackage{graphicx}
\usepackage{esint}

\makeatletter
\@ifundefined{textcolor}{}
{%
 \definecolor{BLACK}{gray}{0}
 \definecolor{WHITE}{gray}{1}
 \definecolor{RED}{rgb}{1,0,0}
 \definecolor{GREEN}{rgb}{0,1,0}
 \definecolor{BLUE}{rgb}{0,0,1}
 \definecolor{CYAN}{cmyk}{1,0,0,0}
 \definecolor{MAGENTA}{cmyk}{0,1,0,0}
 \definecolor{YELLOW}{cmyk}{0,0,1,0}
 }

\makeatother

\usepackage{babel}
\begin{document}
\selectlanguage{american}%

\title{Semiclassical analysis of Bose-Hubbard dynamics}

\selectlanguage{english}%

\author{Hagar Veksler and Shmuel Fishman}

\address{Physics Department, Technion- Israel Institute of Technology, Haifa
3200, Israel}
\selectlanguage{american}%
\begin{abstract}
In this work the two site Bose-Hubbard model is studied analytically
in the limit of weak coupling $u$ and large number of particles $N$.
The semiclassical approximation where $\frac{1}{N}$ plays the role
of Planck's constant was used and perturbation theory to order $u^{2}$
was applied. In particular, the difference in the occupation between
the two sites, where initially all particles are at one site was calculated
analytically. Excellent agreement with the exact numerical solution
was found. This quantity exhibits collapses and revivals that superimpose
rapid oscillations. The occupation difference was calculated also
for the case where initially both sites are occupied provided that
the difference in occupation is sufficiently large. It provides an
analytical description of results that were so far found only numerically.
Similar behavior and analysis are expected for a large variety of
physical situations in optics, atom optics and quantum dynamics of
electrons in Rydberg atoms.
\end{abstract}
\maketitle
\selectlanguage{english}%

\section{introduction}

The physics of Bose-Einstein condensates (BECs) is extensively studied
in the recent years \cite{Pita_book,PS_review,Pethick@Smith}. For
weakly coupled atoms in a large variety of systems the Gross-Pitaevskii
equation (GPE) \cite{Pita_book,Pethick@Smith} describes well the
static properties. For the dynamics, the situation is more complicated
and it is instructive to study simple paradigmatic systems. For the
double well potential, the GPE does not reproduce the correct dynamics.
In particular, the collapse of the amplitude of Rabi oscillations
as a function of time is not reproduced, in contradiction with the
numerical solutions for the many-body system \cite{Milburn,ODell2}.
The double well is a paradigmatic model system that was extensively
studied experimentally \cite{Obert_exp,jeff,shin,dorond}. Much interest
was in the Bosonic Josephson effect \cite{Obert_exp,jeff}. This is
a clear manifestation of macroscopic quantum coherence. It encourages
theoretical exploration of this and related systems \cite{Milburn,Doron,Smerzi,Spekkens&Sipe,ReinhardtPRL}.
If the inter-particle interaction is sufficiently weak so that the
coupling between the lowest levels of the double well and higher levels
can be ignored, the system may be described by the two site Bose Hubbard
(BH) model where bosons can occupy only two sites \cite{Assa}. \foreignlanguage{american}{This
model has been studied numerically and analytically }\cite{Milburn,Doron,Ofir_DW,OfirDW,Obert_the,ODell,Ofir_BH,Kuklov4}\foreignlanguage{american}{.
Fascinating phenomena that were explored are collapse and revival
of the difference in the occupation of the sites} \cite{Milburn}\foreignlanguage{american}{;
and the semiclassically-related statistics of the fluctuations that
are associated with the occupation \cite{dorona,doronb,Doron}. The
latter are important for the fringe visibility in interference experiments
\cite{dorond}}. The understanding of the revivals is crucial for
understanding of the coherence in the system.

Collapses and revivals were observed in experiments where a BEC was
confined to a lattice. The interference pattern of the matter wave
field originating in different lattice sites showed collapses and
revivals as a function of time \cite{Bloch_colapse,Bloch_ex}. These
were found also experimentally for other condensates \cite{Deepak,Deepak2}
and Rydberg atoms \cite{of2,of3}. Coherence was explored for models
of dynamics of atoms on optical lattices in \cite{Fisher3,Fisher4}.
Collapses and revivals were found in numerical calculations following
heuristic arguments and in direct numerical studies of the double
well problem \cite{OfirDW,Ofir_DW,ODell2}. For the two site Bose-Hubbard
model, collapses were found in exact numerical calculations \cite{Ofir_DW,OfirDW}.
Collapses and revivals were found theoretically for interacting bosons
in a harmonic well and the relevant times were estimated \cite{Lewenstein,Wright,wright&walls,castin&dalibard,You}.
These were found also for wave packets in harmonic wells with small
nonlinearities \cite{Mark}. In quantum optics these were found in
the Jaynes-Cummings model \cite{Jaynes_model} analytically and numerically
\cite{Eberly}. It is the closest to the one found in the present
paper for interacting bosons. In optics this phenomenon is well understood
and is known as the Talbot effect \cite{Talbot,Rayleigh} (see also
\cite{carpet,Berry_box}). A related phenomenon is the {}``Quantum
carpet'' \cite{Berry&Sclich,Berry_box}. Collapses and revivals can
be found in many situations. A generic picture is outlined in \cite{of1_pit,PS_review}.

A semiclassical picture for the two site Bose-Hubbard model was developed
and studied in some detail \cite{Obert_the,Milburn,Doron}. The dimensionless
parameter that controls the corresponding classical behavior is
\begin{equation}
u=\frac{UN}{J}\label{eq:u}
\end{equation}
where $U$ is the inter-particle interaction, $J$ is the strength
of the hopping between the sites, while $N$ is the number of particles.
In this picture, $\frac{1}{N}$ plays the role of Planck's constant
and the thermodynamic limit $N\rightarrow\infty$ plays the role of
the classical limit. The various regimes are \cite{Doron}:
\begin{enumerate}
\item Rabi regime $u<1$
\item Josephson regime $1<u<N^{2}$
\item Fock regime $N^{2}<u$.
\end{enumerate}
The Josephson regime is the most extensively studied one \cite{ODell,Obert_the,Doron}.
It exhibits an interesting phase space, with dynamics related to the
experimental observations \cite{Obert_exp,jeff}. We confine ourselves
to the Rabi regime where the classical behavior is very simple. It
enables us to study analytically quantum collapses and revivals that
are crucial in the understanding of coherence.

The present work will follow the formulation presented in detail in
the work of D. Cohen and coworkers \cite{Doron}. In the limit $N\gg1$
we find an analytical formula for the difference between the occupation
of the two sites. It is rare to find such results for interacting
systems and to the best of our knowledge it is the first time such
an analytic expression is found in the present context, namely for
interacting bosons.

The outline of the paper is as follows: In section II the model is
defined and the semiclassical picture is presented. In section III
a transformation to angle action variables is performed and used to
find the energies within the WKB approximation to the order $u^{2}$,
and in section IV it is calculated in standard quantum perturbation
theory. In section V the difference in occupation between the two
sites is calculated for an initial condition where all atoms are on
one site, while in section VI the initial condition where both sites
are occupied is used. The results are discussed in Sec. VII.

\section{The two states Hubbard model}

The two states Hubbard model we study is defined by the Hamiltonian
\begin{equation}
H_{BH}=-J\left(a_{L}^{\dagger}a_{R}+a_{R}^{\dagger}a_{L}\right)+U\left[n_{L}\left(n_{L}-1\right)+n_{R}\left(n_{R}-1\right)\right].\label{eq:H_BH_n}
\end{equation}
The sites are denoted by $L$ (Left) and $R$ (Right). The creation
and annihilation operators on the sites are $a_{L}^{\dagger},a_{R}^{\dagger}$
and $a_{L},a_{R}$ . The number operators for the two sites are $n_{L}=a_{L}^{\dagger}a_{L}$
and $n_{R}=a_{R}^{\dagger}a_{R}$. The commutation relations are $\left[a_{L},a_{L}^{\dagger}\right]=1$,
$\left[a_{R},a_{R}^{\dagger}\right]=1$,\foreignlanguage{american}{
and the units are such that $\hbar=\frac{1}{N}$. It is assumed that
the on site energies on the two sites are identical. The total number
of particles $n_{L}+n_{R}=N$ is conserved. The first term in (\ref{eq:H_BH_n})
represents the hopping between the two sites while the second one
is the energy of the interparticle interaction that in the present
work is assumed to be small. The Hamiltonian (\ref{eq:H_BH_n}) can
be written in the form
\begin{equation}
H_{BH}=-J\left(a_{L}^{\dagger}a_{R}+a_{R}^{\dagger}a_{L}\right)+U\left(a_{R}^{\dagger}a_{R}^{\dagger}a_{R}a_{R}+a_{L}^{\dagger}a_{L}^{\dagger}a_{L}a_{L}\right).\label{eq:H_BH}
\end{equation}
 By using the angular momentum operators (see for example \cite{Doron})
\begin{equation}
\begin{array}{ccl}
S_{x} & = & \frac{1}{2N}\left(a_{R}^{\dagger}a_{L}+a_{L}^{\dagger}a_{R}\right)\\
S_{y} & = & \frac{i}{2N}\left(a_{R}^{\dagger}a_{L}-a_{L}^{\dagger}a_{R}\right)\\
S_{z} & = & \frac{1}{2N}\left(a_{L}^{\dagger}a_{L}-a_{R}^{\dagger}a_{R}\right)=\frac{1}{2N}\left(n_{L}-n_{R}\right)
\end{array},\label{eq:S_def}
\end{equation}
the Hamiltonian (\ref{eq:H_BH}) can be written up to a constant as
(shown in App. A)
\begin{equation}
H'=-2JNS_{x}+2UN^{2}S_{z}^{2}.\label{eq:H_s}
\end{equation}
Namely,
\begin{equation}
H_{BH}=H'+C_{N}
\end{equation}
where $C_{N}=\frac{1}{2}N^{2}U-NU$. The operators (\ref{eq:S_def})
satisfy the commutation relations of angular momentum operators
\begin{equation}
\begin{array}{ccc}
\left[S_{x},S_{y}\right] & = & \frac{i}{N}S_{z}\\
\left[S_{y},S_{z}\right] & = & \frac{i}{N}S_{x}\\
\left[S_{z},S_{x}\right] & = & \frac{i}{N}S_{y}
\end{array}\label{eq:com}
\end{equation}
as can be easily verified. These are the standard commutation relations
of the angular momentum operators. It is convenient to measure the
energy in units of $2JN$, and work with the Hamiltonian 
\begin{equation}
H=-S_{x}+uS_{z}^{2}\label{eq:H_sx_sz}
\end{equation}
where $u\equiv\frac{UN}{J}$ (see (\ref{eq:H_BH_n})). Since (\ref{eq:S_def})
are angular momentum operators, and the eigenvalues of $NS_{z}$ are
integers $n$ satisfying $-\frac{N}{2}<n<\frac{N}{2}$, $S^{2}=S_{x}^{2}+S_{y}^{2}+S_{z}^{2}=\frac{1}{2N}\left(\frac{N}{2}+1\right)$.
For large $N$ the semi-classical limit is justified. In the classical
limit, the equation of motion can be obtained by replacing $\frac{N}{i}\left[f,g\right]\longrightarrow\left\{ f,g\right\} $
where $\left\{ f,g\right\} $ are the Poisson's brackets. These are
the Hamilton equations obtained from (\ref{eq:H_sx_sz}). As the total
number of particles $N$ is conserved, the total angular momentum
$S^{2}$ is conserved as well. Therefore, the vector $\overrightarrow{S}=\left(S_{x},S_{y},S_{z}\right)$
lies on the Bloch sphere of radius $\frac{1}{2}$ and it is possible
to write
\begin{equation}
\begin{array}{ccc}
S_{y} & = & \sqrt{\frac{1}{4}-S_{x}^{2}}\cos\varphi\\
S_{z} & = & \sqrt{\frac{1}{4}-S_{x}^{2}}\sin\varphi
\end{array}
\end{equation}
where $0<\varphi<2\pi$ is an angle circling the $S_{x}$ axis. Now,
the Hamiltonian (\ref{eq:H_sx_sz}) takes the form
\begin{equation}
H=-S_{x}+u\left(\frac{1}{4}-S_{x}^{2}\right)\sin^{2}\varphi.\label{eq:H_sx_fi}
\end{equation}
}

\section{The semi-classical calculation of the spectrum}

\selectlanguage{american}%
In the absence of inter-particle interactions $\left(u=0\right)$,
the bosons undergo Rabi oscillations and the phase space trajectories
circle around the $S_{x}$ axis with frequency $2J$. We would like
to study the dynamics in the Rabi regime $u\ll1$ (weak inter-particle
interactions) by using semi-classical methods. Our aim is to find
the spectrum of (\ref{eq:H_sx_fi}) by using WKB quantization for
the action variable (a similar approach was adopted by \cite{Doron}
for the Josephson regime $1<u<N^{2}$). $S_{x}$ and $\varphi$ are
canonically conjugate variables. Their variation is given by Hamilton's
equations generated by $H$ of (\ref{eq:H_sx_sz}). It was verified
that these are identical to the equations satisfied by the components
of $\vec{S}$. We turn now to calculate the action variable via \cite{Tabor}
\begin{equation}
I=\frac{1}{2\pi}\int_{0}^{2\pi}S_{x}d\varphi.\label{eq:I_int}
\end{equation}
For this we use the relation between $S_{x}$ and $\varphi$ given
by 
\begin{equation}
S_{x}=\frac{-1\pm\sqrt{1+u\sin^{2}\varphi\left(u\sin^{2}\varphi-4H\right)}}{2u\sin^{2}\varphi}\label{eq:Sx}
\end{equation}
where only the $+$ solution is consistent with (\ref{eq:H_sx_fi})
for $u=0$. In the first order in $u$, the action can be calculated
from 
\begin{equation}
S_{x}\approx-H+\frac{1}{4}u\sin^{2}\varphi-uH^{2}\sin^{2}\varphi\label{eq:Sx_first}
\end{equation}
and by (\ref{eq:I_int}),
\begin{equation}
I\approx-H+\frac{1}{8}u-\frac{1}{2}uH^{2}.\label{eq:I_int-1}
\end{equation}
Now, one can write the Hamiltonian in terms of $I$ as
\begin{equation}
H\approx\frac{-1\pm\sqrt{1+\frac{1}{4}u^{2}-2uI}}{u}
\end{equation}
where only the $+$ solution satisfies (\ref{eq:I_int-1}) for $u=0$.
Therefore, to the first order in $u$,
\begin{equation}
H\approx-I+\frac{1}{8}u-\frac{1}{2}uI^{2}.\label{eq:H_I}
\end{equation}
The action variable is quantized \cite{Tabor} so that
\begin{equation}
I_{n}=\frac{n}{N}
\end{equation}
where $n=-\frac{N}{2},...,\frac{N}{2}$ are integers. Note that $\dot{\varphi}=-\frac{\partial H}{\partial S_{x}}\approx-1$
for small $u$ and therefore $\dot{\varphi}$ never vanishes. Consequently
the Maslov index vanishes. Hence, the spectrum of the Hamiltonian
(\ref{eq:H_sx_sz}) is
\begin{equation}
E_{n}^{\left(1\right)}\approx-\frac{n}{N}+\frac{1}{8}u-\frac{1}{2N^{2}}un^{2}.\label{eq:En}
\end{equation}
In order to compare the energies (\ref{eq:En}) to the exact spectrum
of the BH model (\ref{eq:H_BH}) (which can be obtained by diagonalizing
the Hamiltonian matrix), we should multiply it by $2JN$ and add the
constants which were omitted in (\ref{eq:H_s}) (see (\ref{eq:constants}))
, namely,
\begin{equation}
E_{n}^{\left(BH1\right)}=2JNE_{n}^{\left(1\right)}+C_{N}\approx2J\left(-n+\frac{3}{8}uN-\frac{1}{2}u-\frac{1}{2N}un^{2}\right).\label{eq:En-1}
\end{equation}
For $u<1$, This spectrum is a good approximation to the exact BH
spectrum (see Fig. 1).

\selectlanguage{english}%
In second order in $u$, one finds:

\begin{equation}
\begin{array}{ccl}
S_{x} & \approx & -H+\frac{1}{4}u\sin^{2}\varphi-uH^{2}\sin^{2}\varphi+\frac{1}{2}u^{2}H\sin^{4}\varphi-2u^{2}H^{3}\sin^{4}\varphi\end{array},\label{eq:Sx_sec}
\end{equation}
which leads to an action variable of the form
\begin{equation}
\begin{array}{ccl}
I & \approx & -H+\frac{1}{8}u-\frac{1}{2}uH^{2}+\frac{3}{16}u^{2}H-\frac{3}{4}u^{2}H^{3}\end{array}.\label{eq:IH3}
\end{equation}
In order to find the corrections to the spectrum (\ref{eq:En}), we
substitute $H=E_{n}^{\left(1\right)}+u^{2}\cdot\delta_{n}$ in (\ref{eq:IH3})
and keep terms up to the second order in $u$, resulting in
\begin{equation}
\delta_{n}=-\frac{n}{16N}+\frac{n^{3}}{4N^{3}}
\end{equation}
and
\begin{equation}
E_{n}^{\left(2\right)}\approx-\frac{n}{N}+\frac{1}{8}u-\frac{1}{2N^{2}}un^{2}-\frac{nu^{2}}{16N}+\frac{n^{3}u^{2}}{4N^{3}}\label{eq:E_BH2-1}
\end{equation}
leading to
\begin{equation}
E_{n}^{\left(BH2\right)}\approx2J\left(-n+\frac{3}{8}uN-\frac{1}{2}u-\frac{1}{2N}un^{2}-\frac{1}{16}u^{2}n+\frac{1}{4N^{2}}u^{2}n^{3}\right).\label{eq:E_BH2}
\end{equation}
Numerical calculations (Fig. 1) verify that the spectrum $E_{n}^{\left(BH2\right)}$
is indeed closer to the BH spectrum than $E_{n}^{\left(BH1\right)}$.
Although the second order correction is extremely small compared to
the first order, it turns out to be of great importance for the dynamics
and in particular for the shape of the revival peaks as will be shown
in Sec. V. Note that the agreement is very good even for $u$ that
is not much smaller than $1$. There are predictions based on low
order perturbation theory that hold even when the perturbations are
not very small \cite{Doron_cp1,doron_cp2,Doron_cp3,Doron_cp4}. For
the present work, it is particularly instructive to note Eq. (A.2)
and (A.3) of Ref. \cite{Doron_cp4}. The result of (\ref{eq:E_BH2})
is actually not the one of quantum perturbation theory but the expansion
to the order $u^{2}$ of the leading semiclassical result. This expansion
is convergent in general for $u<\frac{1}{2}$, and for small $n$
used in the paper, it is sufficient that $u<1$ as can be seen from
(\ref{eq:Sx}). In the next section, standard quantum perturbation
theory is used and we note that it requires $uN<1$.

\selectlanguage{american}%
\begin{figure}[H]
\selectlanguage{english}%
\includegraphics[scale=0.5]{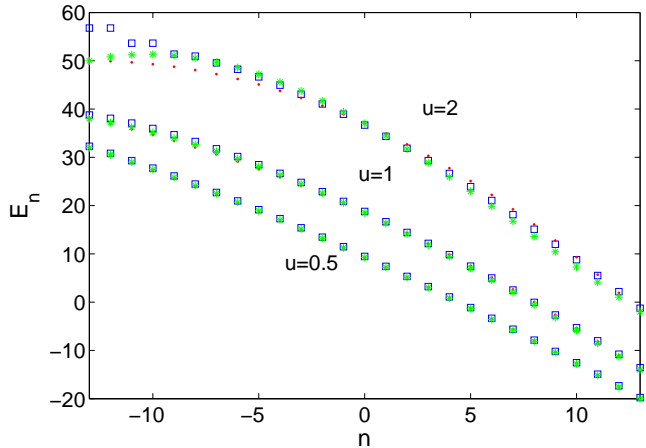}

\selectlanguage{american}%
\caption{\selectlanguage{english}%
(Color online) The energy spectrum of the BH Hamiltonian for $J=1$
and $N=26$. The blue squares are obtained by numerical diagonalization
of the Hamiltonian matrix for the Hamiltonian (\ref{eq:H_BH}). The
red dots are analytically calculated to the first order in $u$ (\ref{eq:En-1})
and the green stars are analytically calculated to the second order
in $u$ (see (\ref{eq:E_BH2})). \selectlanguage{american}
}
\end{figure}
A natural question is what are the corrections to the leading order
in the Semiclassical expansion presented here. In App. D it is shown
that the correction is of order $\frac{1}{N^{2}}$. Hence it is of
the form $\frac{1}{N^{2}}f\left(H\right)$. Then, this term should
be added to the RHS of (\ref{eq:IH3}) leading to an additional correction
to the energy levels. In App. D. we estimate this correction for representative
values of the parameters and find it to be extremely small.

\selectlanguage{english}%

\section{The pertubative calculation of the spectrum}

It is possible to calculate the spectrum of (\ref{eq:H_sx_sz}) by
using standard quantum perturbation theory for small $u$. The perturbation
series is likely to converge if $uN<1$ since the energy differences
are of order $\frac{1}{N}$, see (\ref{eq:dE}). In the first order
in $u$,
\begin{equation}
\widetilde{E}^{\left(1\right)}=-\frac{n}{N}+u\left\langle n\left|S_{z}^{2}\right|n\right\rangle .\label{eq:first_E}
\end{equation}
The matrix element $\left\langle k\left|S_{z}^{2}\right|n\right\rangle $
can be calculated easily by using the relation
\begin{equation}
S_{z}=\frac{1}{2}\left(\widetilde{S}_{+}+\widetilde{S}_{-}\right)
\end{equation}
where $\widetilde{S}_{\pm}=\left(S_{z}\mp iS_{y}\right)$ given by
(\ref{eq:S_def}) are ladder operators satisfying 
\begin{equation}
\widetilde{S}_{\pm}\left|n\right\rangle =\frac{1}{N}\sqrt{\left(\frac{N}{2}\pm n+1\right)\left(\frac{N}{2}\mp n\right)}\left|n\pm1\right\rangle 
\end{equation}
\begin{equation}
\widetilde{S}_{\pm}\left|n\right\rangle =\frac{1}{2}\sqrt{1+\frac{2}{N}-\frac{4n^{2}}{N^{2}}\mp\frac{4n}{N^{2}}}\left|n\pm1\right\rangle .
\end{equation}
Hence,
\begin{equation}
\begin{array}{ccl}
S_{z}^{2}\left|n\right\rangle  & = & \frac{1}{4}\left(\widetilde{S}_{+}^{2}+\widetilde{S}_{-}^{2}+\widetilde{S}_{+}\widetilde{S}_{-}+\widetilde{S}_{-}\widetilde{S}_{+}\right)\left|n\right\rangle \\
 & = & \frac{1}{16}\cdot\sqrt{\left(1+\frac{2}{N}-\frac{4n^{2}}{N^{2}}-\frac{4n}{N^{2}}\right)\cdot\left(1+\frac{2}{N}-\frac{4\left(n+1\right)^{2}}{N^{2}}-\frac{4\left(n+1\right)}{N^{2}}\right)}\left|n+2\right\rangle \\
 &  & +\frac{1}{16}\cdot\sqrt{\left(1+\frac{2}{N}-\frac{4n^{2}}{N^{2}}+\frac{4n}{N^{2}}\right)\cdot\left(1+\frac{2}{N}-\frac{4\left(n-1\right)^{2}}{N^{2}}+\frac{4\left(n-1\right)}{N^{2}}\right)}\left|n-2\right\rangle \\
 &  & +\frac{1}{16}\cdot\sqrt{\left(1+\frac{2}{N}-\frac{4n^{2}}{N^{2}}+\frac{4n}{N^{2}}\right)\cdot\left(1+\frac{2}{N}-\frac{4\left(n-1\right)^{2}}{N^{2}}-\frac{4\left(n-1\right)}{N^{2}}\right)}\left|n\right\rangle \\
 &  & +\frac{1}{16}\cdot\sqrt{\left(1+\frac{2}{N}-\frac{4n^{2}}{N^{2}}-\frac{4n}{N^{2}}\right)\cdot\left(1+\frac{2}{N}-\frac{4\left(n+1\right)^{2}}{N^{2}}+\frac{4\left(n+1\right)}{N^{2}}\right)}\left|n\right\rangle 
\end{array}.\label{eq:Sz_M}
\end{equation}
Assuming $N\gg1$, we expand $\left\langle n\left|S_{z}^{2}\right|n\right\rangle $
to the second order in $\frac{1}{N}$ and get
\begin{equation}
\begin{array}{ccl}
\left\langle n\left|S_{z}^{2}\right|n\right\rangle  & \approx & \frac{1}{16}\cdot\sqrt{1+\frac{4}{N}-\frac{8n^{2}}{N^{2}}+\frac{8n}{N^{2}}+\frac{4}{N^{2}}}\\
 &  & +\frac{1}{16}\cdot\sqrt{1+\frac{4}{N}-\frac{8n^{2}}{N^{2}}-\frac{8n}{N^{2}}+\frac{4}{N^{2}}}\\
 & \approx & \frac{1}{16}\left[2+\frac{4}{N}-\frac{8n^{2}}{N^{2}}\right]
\end{array}.\label{eq:Sz_M-1}
\end{equation}
resulting in
\begin{equation}
\widetilde{E}^{\left(1\right)}=-\frac{n}{N}+\frac{u}{8}\left[1+\frac{2}{N}-\frac{4n^{2}}{N^{2}}\right],\label{eq:first_E-1}
\end{equation}
which is equivalent to the semiclassical correction calculated in
(\ref{eq:En}), if $\frac{1}{N}$ is ignored compared to $1$.

The energies to the second order in $u$ are
\begin{equation}
\widetilde{E}^{\left(2\right)}=-\frac{n}{N}+u\left\langle n\left|S_{z}^{2}\right|n\right\rangle +u^{2}\sum_{k\neq n}\frac{\left|\left\langle k\left|S_{z}^{2}\right|n\right\rangle \right|^{2}}{E_{n}^{\left(0\right)}-E_{k}^{\left(0\right)}}.\label{eq:second_E-2}
\end{equation}
The energy differences are
\begin{equation}
\widetilde{E}_{n}^{\left(0\right)}-\widetilde{E}_{k}^{\left(0\right)}=\frac{k-n}{N}\label{eq:dE}
\end{equation}
and the matrix elements $\left\langle k\left|S_{z}^{2}\right|n\right\rangle $
does not vanish only for $k=n\pm2$. Therefore, 
\begin{equation}
\begin{array}{ccl}
\sum_{k\neq n}\frac{\left|\left\langle k\left|S_{z}^{2}\right|n\right\rangle \right|^{2}}{E_{n}^{\left(0\right)}-E_{k}^{\left(0\right)}} & = & \frac{N}{2\cdot16^{2}}\left(1+\frac{2}{N}-\frac{4n^{2}}{N^{2}}-\frac{4n}{N^{2}}\right)\cdot\left(1+\frac{2}{N}-\frac{4n^{2}}{N^{2}}-\frac{12n}{N^{2}}-\frac{8}{N^{2}}\right)\\
 &  & -\frac{N}{2\cdot16^{2}}\left(1+\frac{2}{N}-\frac{4n^{2}}{N^{2}}+\frac{4n}{N^{2}}\right)\cdot\left(1+\frac{2}{N}-\frac{4n^{2}}{N^{2}}+\frac{12n}{N^{2}}-\frac{8}{N^{2}}\right)\\
 & = & \frac{1}{16^{2}}\left[-\frac{4n}{N}\left(1+\frac{2}{N}-\frac{4\left(n^{2}+2\right)}{N^{2}}\right)-\frac{12n}{N}\left(1+\frac{2}{N}-\frac{4n^{2}}{N^{2}}\right)\right]\\
 & = & -\frac{4n}{16^{2}N}\left[4+\frac{8}{N}-\frac{8}{N^{2}}-\frac{16n^{2}}{N^{2}}\right]=-\frac{n}{16N}\left[1+\frac{2}{N}-\frac{2}{N^{2}}-\frac{4n^{2}}{N^{2}}\right]
\end{array}\label{eq:Sz2_matrix-1}
\end{equation}
and,
\begin{equation}
\widetilde{E}^{\left(2\right)}=-\frac{n}{N}+\frac{u}{8}\left[1+\frac{2}{N}-\frac{4n^{2}}{N^{2}}\right]+\frac{u^{2}}{16}\left[-\frac{n}{N}\left(1+\frac{2}{N}-\frac{2}{N^{2}}\right)+\frac{4n^{3}}{N^{3}}\right]
\end{equation}
which is equivalent to the semiclassical correction calculated in
(\ref{eq:E_BH2-1}), when $\frac{1}{N}$ is ignored compared to $1$.

\section{Dynamics}

In this section, our aim is to derive an analytic expression for the
expectation value of $S_{z}\left(t\right)$ that is the difference
in occupation of the two sites where the initial condition is that
all the bosons occupy the state $L$ and $\left\langle S_{z}\right\rangle =\frac{1}{2}$
(it is the north pole of the phase space Bloch sphere). In the framework
of the BH model (\ref{eq:H_BH}), it is possible to calculate $S_{z}\left(t\right)$
numerically \cite{Milburn}. The resulting $S_{z}\left(t\right)$
is a series of collapses and revivals, superimposed on rapid oscillations.
We would like to utilize the spectrum (\ref{eq:E_BH2}) in order to
study analytically the dynamics in the Rabi regime $u<1$. 

In absence of inter-particle interactions ($u=0$), the operator $S_{x}$
commutes with the Hamiltonian (\ref{eq:H_sx_sz}). Hence, for $u<1$,
the eigenstates of (\ref{eq:H_sx_sz}) can be approximated by the
eigenstates of $S_{x}$ (corrections of higher order will be discussed
later), namely by
\begin{equation}
\left|n\right\rangle \equiv\frac{1}{\sqrt{\left(\frac{N}{2}+n\right)!\left(\frac{N}{2}-n\right)!}}\left(a_{+}^{\dagger}\right)^{\frac{N}{2}+n}\left(a_{-}^{\dagger}\right)^{\frac{N}{2}-n}\left|0\right\rangle .\label{eq:S_x_eigen}
\end{equation}
where $a_{\pm}^{\dagger}=\frac{1}{\sqrt{2}}\left(a_{L}^{\dagger}\pm a_{R}^{\dagger}\right)$
and 
\begin{equation}
\left[a_{+},a_{-}\right]=0\label{eq:a+-}
\end{equation}
 
\begin{equation}
\left[a_{+},a_{+}^{\dagger}\right]=1\label{eq:a++}
\end{equation}
\begin{equation}
\left[a_{-},a_{-}^{\dagger}\right]=1\label{eq:a--}
\end{equation}
The reason for (\ref{eq:S_x_eigen}) is that
\begin{equation}
S_{x}=\frac{1}{2N}\left(a_{+}^{\dagger}a_{+}-a_{-}^{\dagger}a_{-}\right).
\end{equation}
In what follows, we calculate the evolution of the operator 
\begin{equation}
\widetilde{S}_{+}\equiv N\left(S_{z}-iS_{y}\right)=\frac{1}{2}\left(a_{L}^{\dagger}+a_{R}^{\dagger}\right)\left(a_{L}-a_{R}\right)=a_{+}^{\dagger}a_{-}\label{eq:sy+sz}
\end{equation}
for the initial condition 
\begin{equation}
\left|\psi\left(t=0\right)\right\rangle =\frac{1}{\sqrt{N!}}\left(a_{L}^{\dagger}\right)^{N}=\frac{1}{2^{N/2}\sqrt{N!}}\left(a_{+}^{\dagger}+a_{-}^{\dagger}\right)^{N}.
\end{equation}
It is useful to expand $\left|\psi\left(t=0\right)\right\rangle $
in the basis of (\ref{eq:S_x_eigen}),
\begin{equation}
\left|\psi\left(t=0\right)\right\rangle =\sum_{n=-N/2}^{N/2}c_{n}\left|n\right\rangle ,\label{eq:i}
\end{equation}
where
\begin{equation}
c_{n}=\frac{1}{2^{N/2}\sqrt{N!}}\cdot\left(\begin{array}{c}
N\\
\frac{N}{2}+n
\end{array}\right)\cdot\sqrt{\left(\frac{N}{2}+n\right)!\left(\frac{N}{2}-n\right)!}=\frac{1}{2^{N/2}}\sqrt{\left(\begin{array}{c}
N\\
\frac{N}{2}+n
\end{array}\right)}.
\end{equation}
For $N\gg1$ and $n\ll\frac{N}{2}$, the binomial coefficients can
be approximated by a Gaussian, 
\begin{equation}
c_{n}\approx\left(\frac{2}{\pi N}\right)^{\frac{1}{4}}e^{-\frac{n^{2}}{N}}.\label{eq:cn}
\end{equation}
We note that the normalized difference between the occupation of the
two sites is 
\begin{equation}
\Delta\left(t\right)=\left\langle \psi\left|S_{z}\right|\psi\right\rangle =\frac{1}{N}\mathrm{Re}\left\langle \psi\left|\widetilde{S}_{+}\right|\psi\right\rangle \label{eq:Delta}
\end{equation}
where $\widetilde{S}_{+}\equiv N\left(S_{z}-iS_{y}\right)=\frac{1}{2}\left(a_{L}^{\dagger}+a_{R}^{\dagger}\right)\left(a_{L}-a_{R}\right)=a_{+}^{\dagger}a_{-}$.
In the basis $\left\{ \left|n\right\rangle \right\} $, $\widetilde{S}_{+}$
is a raising operator, therefore, 
\begin{equation}
\widetilde{S}_{+}\left|n\right\rangle =\sqrt{\left(\frac{N}{2}+n+1\right)\left(\frac{N}{2}-n\right)}\left|n+1\right\rangle ,\label{eq:s+}
\end{equation}
we used (\ref{eq:a+-}), (\ref{eq:a++}) and (\ref{eq:a--}).The expectation
value of $\widetilde{S}_{+}$ at time $t$ for the initial condition
(\ref{eq:i}) is
\begin{equation}
\left\langle \psi\left(t\right)\left|\widetilde{S}_{+}\right|\psi\left(t\right)\right\rangle =\sum_{n=-N/2}^{N/2}\sqrt{\left(\frac{N}{2}+n+1\right)\left(\frac{N}{2}-n\right)}c_{n}c_{n+1}e^{-i\left(E_{n}^{\left(BH2\right)}-E_{n+1}^{\left(BH2\right)}\right)t}.\label{eq:usm}
\end{equation}
First, note that
\begin{equation}
c_{n}c_{n+1}=\frac{\sqrt{2}}{\sqrt{\pi N}}e^{-\frac{2n^{2}+2n+1}{N}}.
\end{equation}
According to (\ref{eq:E_BH2}),
\begin{equation}
E_{n}^{\left(BH2\right)}-E_{n+1}^{\left(BH2\right)}=J\left(2+\frac{2}{N}un+\frac{1}{8}u^{2}-\frac{3}{2N^{2}}u^{2}n^{2}+\frac{u}{N}-\frac{3}{2N^{2}}u^{2}n-\frac{1}{2N^{2}}u^{2}\right).\label{eq:Spect_diff2}
\end{equation}
Since $u<1$, we can neglect the term $\frac{3}{2N^{2}}u^{2}n$ which
is much smaller than $\frac{2}{N}un$. For large $N$ and $n\ll\frac{N}{2}$,
(\ref{eq:usm}) can be written in the form
\begin{equation}
\left\langle \psi\left(t\right)\left|\widetilde{S}_{+}\right|\psi\left(t\right)\right\rangle =\frac{N}{2}\widetilde{S}e^{-i\phi t}
\end{equation}
where
\begin{equation}
\phi=J\left(2+\frac{1}{8}u^{2}+\frac{u}{N}-\frac{1}{2N^{2}}u^{2}\right)\label{eq:full_fi}
\end{equation}
and 
\begin{equation}
\widetilde{S}=\frac{\sqrt{2}}{\sqrt{\pi N}}\sum_{n=-N/2}^{N/2}e^{-\frac{2n^{2}+2n+1}{N}}e^{-i\frac{J}{N}\left(2un-\frac{3}{2N}u^{2}n^{2}-\frac{3}{2N}u^{2}n\right)t}.
\end{equation}
We note that $e^{-i\phi t}$ is a rapidly oscillating function of
$t$ with a period that is approximately $\frac{\pi}{J}$. We turn
now to explore the envelope of $\widetilde{S}$. Since $n$ is an
integer, in first order in $u$, the envelope of the sum (\ref{eq:usm})
is a periodic function of $t$ with period (revival time) of 
\begin{equation}
T_{R}=\frac{\pi N}{uJ}.\label{eq:Tr}
\end{equation}
Actually $T_{R}$ is the inverse of the coefficient of the linear
term in $n$ in the RHS of (\ref{eq:Spect_diff2}), namely $\frac{1}{T_{R}}=\frac{J}{2\pi}\left(\frac{2}{N}u+\frac{3u^{2}}{2N^{2}}\right)$.
The estimate (\ref{eq:Tr}) assumes $\frac{u}{N}\ll1$. The terms
proportional to $u^{2}$ in (\ref{eq:Spect_diff2}) are ignored for
the same reason, taking into account that in what follows only terms
where $n\ll N$ are important. Around the $m$-th revival, we write
$t=m\cdot T_{R}+\tau$ with $-\frac{1}{2}T_{R}<\tau<\frac{1}{2}T_{R}$
and write $\widetilde{S}=\sum_{m}\widetilde{S}_{m}$ where
\begin{equation}
\begin{array}{ccc}
\widetilde{S}_{m} & = & \frac{\sqrt{2}}{\sqrt{\pi N}}\sum_{-\infty}^{\infty}e^{-\left(2n^{2}+2n+1\right)/N}e^{-i\frac{J}{N}\left(2un-\frac{3}{2N}u^{2}n^{2}\right)\cdot\left(m\cdot T_{R}+\tau\right)}\\
 & = & \frac{\sqrt{2}}{\sqrt{\pi N}}\sum_{-\infty}^{\infty}e^{-\left(2n^{2}+2n+1\right)/N}e^{+i\frac{3}{2N^{2}}u^{2}n^{2}J\cdot\left(m\cdot T_{R}+\tau\right)-\frac{2}{N}iJun\tau}
\end{array}.\label{eq:usm-1-2}
\end{equation}
We approximate the sum by an integral
\begin{equation}
\widetilde{S}_{m}\approx\frac{\sqrt{2}}{\sqrt{\pi N}}\int_{-\infty}^{\infty}e^{-\left(2n^{2}+2n+1\right)/N}e^{+i\frac{3}{2N^{2}}u^{2}n^{2}J\cdot\left(m\cdot T_{R}+\tau\right)-\frac{2}{N}iJun\tau}dn.\label{eq:Sm_int}
\end{equation}
What enables to approximate the sum over $n$ by an integral is the
fact that in the vicinity of a revival $\frac{J}{N}u\tau+\frac{3}{N^{2}}u^{2}n\tau$
is small (while $\frac{J}{N}unt$ is typically large). The integral
was calculated in App. B (where we should take $\beta=\gamma=1$ and
$\bar{m}=m$ for the calculations of the present section) using $\psi$
in the order $u^{0}$ and in App. C the corrections of the order $u$
and $u^{2}$ were added. In App. E it is verified that the semiclassical
wave function gives the same result. The result is
\begin{equation}
\widetilde{S}_{m}=\frac{\sqrt{2}R}{D_{D}^{1/4}}e^{\frac{D_{R}}{D_{D}}-\frac{1}{N}+i\left(\phi_{s}+\phi'_{s}\right)}\label{eq:Sm}
\end{equation}
where $D_{R},D_{D}$ and $\phi_{s}$ are given by (\ref{eq:DR}),
(\ref{eq:DDD}), (\ref{eq:fS}),
\begin{equation}
\frac{D_{R}}{D_{D}}\approx\frac{-2J^{2}u^{2}\left(\tau+\frac{3m\pi}{2J}\right)^{2}+2+\frac{9}{2}u^{2}m^{2}\pi^{2}}{N\left[4+\frac{9}{4}u^{2}\left(m\pi+\frac{J}{N}u\tau\right)^{2}\right]},\label{eq:DRDD}
\end{equation}
 and $R$ and $\phi'_{s}$ will be calculated in what follows.
\begin{equation}
R=\left|1+\frac{u}{4}\left(\frac{B}{A}-1\right)-\frac{u^{2}}{32}\right|\approx1-\frac{u^{2}}{32}\approx e^{-\frac{u^{2}}{32}}\label{eq:AB1}
\end{equation}
and
\begin{equation}
\tan\phi'_{s}\approx\frac{u^{2}}{8}\left(2J\tau+\frac{3}{2}m\pi\right).\label{eq:ABf}
\end{equation}
For small $u$, $\phi'_{s}\approx\tan\phi'_{s}\approx\frac{u^{2}}{8}\left(2J\tau+\frac{3}{2}m\pi\right)$.

The resulting $\widetilde{S}_{m}$ (\ref{eq:Sm}) is approximately
a Gaussian of a width
\begin{equation}
\Delta t_{R}^{m}\approx\frac{\sqrt{2N\left(1+\frac{9}{16}u^{2}m^{2}\pi^{2}\right)}}{Ju}.\label{eq:dtRm}
\end{equation}
Therefore, $\tau\lesssim\frac{\sqrt{2N}}{Ju}$ and $\frac{J}{N}u\tau$
(see denominator on (\ref{eq:DRDD})) is of order $\frac{1}{\sqrt{N}}$
and can be neglected compared to $m\pi$, as was done in (\ref{eq:dtRm})
and in the following equations.

For small $m$, $\Delta t_{R}^{m}\ll T_{R}$. However, there is an
$m_{max}$ where the width $\Delta t_{R}^{m}$ is comparable to $T_{R}$
and then the revivals mix and our calculations are not valid. Defining
$m_{max}$ by $\Delta t_{R}^{m_{max}}=\frac{1}{2}T_{R}$, we estimate
\begin{equation}
m_{max}=\frac{\sqrt{2\left(\pi^{2}N-8\right)}}{3u\pi},\label{eq:m_max}
\end{equation}
namely, the revivals start to mix at time 
\begin{equation}
T_{B}=m_{max}T_{R}\approx\frac{\pi\sqrt{2}N^{\frac{3}{2}}}{3u^{2}J}.\label{eq:wash_time}
\end{equation}
For times $t<m_{max}T_{R}$, in the leading order in $u$, $\widetilde{S}$
can be approximated by 
\begin{equation}
\widetilde{S}=\sum_{m}\frac{e^{-\frac{u^{2}}{32}}}{\left[1+\frac{9}{16}u^{2}m^{2}\pi^{2}\right]^{1/4}}\exp\left[\frac{-\frac{1}{2}J^{2}u^{2}\left(t+\frac{3m\pi}{2J}-mT_{R}\right)^{2}+\eta}{N\left(1+\frac{9}{16}u^{2}m^{2}\pi^{2}\right)}+i\left(\phi_{s}+\phi'_{s}\right)\right].
\end{equation}
Therefore,
\begin{equation}
\left\langle \psi\left(t\right)\left|\widetilde{S}_{+}\right|\psi\left(t\right)\right\rangle =\frac{N}{2}\sum_{m}\frac{e^{-\frac{u^{2}}{32}}}{\left[1+\frac{9}{16}u^{2}m^{2}\pi^{2}\right]^{1/4}}\exp\left[\frac{-\frac{1}{2}J^{2}u^{2}\left(t+\frac{3m\pi}{2J}-mT_{R}\right)^{2}+\eta}{N\left(1+\frac{9}{16}u^{2}m^{2}\pi^{2}\right)}+i\left(\phi_{1}-\phi t\right)\right]
\end{equation}
and (see (\ref{eq:Delta}))
\begin{equation}
\Delta\left(t\right)=\frac{1}{2}\sum_{m}\frac{e^{-\frac{u^{2}}{32}}}{\left[1+\frac{9}{16}u^{2}m^{2}\pi^{2}\right]^{1/4}}\exp\left[\frac{-\frac{1}{2}J^{2}u^{2}\left(t+\frac{3m\pi}{2J}-mT_{R}\right)^{2}+\eta}{N\left(1+\frac{9}{16}u^{2}m^{2}\pi^{2}\right)}\right]\cos\left(\phi_{1}-\phi t\right),\label{eq:S_z_good}
\end{equation}
\begin{equation}
\left\langle \psi\left(t\right)\left|\widetilde{S}_{y}\right|\psi\left(t\right)\right\rangle =\frac{1}{2}\sum_{m}\frac{e^{-\frac{u^{2}}{32}}}{\left[1+\frac{9}{16}u^{2}m^{2}\pi^{2}\right]^{1/4}}\exp\left[\frac{-\frac{1}{2}J^{2}u^{2}\left(t+\frac{3m\pi}{2J}-mT_{R}\right)^{2}+\eta}{N\left(1+\frac{9}{16}u^{2}m^{2}\pi^{2}\right)}\right]\sin\left(\phi t-\phi_{1}\right).
\end{equation}
where
\begin{equation}
\eta=\frac{1}{2}+\frac{9}{8}u^{2}m^{2}\pi^{2}\label{eq:L}
\end{equation}
and
\begin{equation}
\phi\approx J\left(2+\frac{1}{8}u^{2}+\frac{u}{N}\right).\label{eq:fi_main}
\end{equation}
In the expression for $\phi$ we neglected $\frac{1}{N^{2}}$ compared
to $1$ in (\ref{eq:full_fi}). The other phase variable is 
\begin{equation}
\phi_{1}=\phi_{s}+\phi'_{s}\approx\frac{u^{2}}{8}\left(2J\tau+\frac{3}{2}m\pi\right)+u\left(\frac{J\tau}{N}+\frac{3}{8}\left(1+\frac{1}{N}\right)\left(m\cdot\pi+\frac{J}{N}u\tau\right)\right),\label{eq:full_fi1}
\end{equation}
neglecting $\frac{1}{N}$ compared to $1$, one finds
\begin{equation}
\phi_{1}=\phi_{s}+\phi'_{s}\approx\frac{u^{2}}{8}\left(2J\tau+\frac{3}{2}m\pi\right)+u\left(\frac{J\tau}{N}+\frac{3}{8}\left(m\cdot\pi+\frac{J}{N}u\tau\right)\right).\label{eq:fi1_main}
\end{equation}
The evolution of the expectation of the normalized difference in occupation
of the two sites is the main result of the present work. In Fig. 2
it is compared to exact results found by numerical diagonalization
of the Hamiltonian (\ref{eq:H_BH}), for $u=\frac{1}{2},N=100$ and
$J=1$. In Fig. 2 as well as in Figs. 3 the expressions (\ref{eq:fi_main})
and (\ref{eq:fi1_main}) for the phases were used. We checked that
if (\ref{eq:full_fi}) and (\ref{eq:full_fi1}) are used instead,
the results cannot be distinguished in the plots. We note remarkable
agreement of the envelope with the exact numerical result. The rapid
oscillations, exhibit good agreement for short times (Fig. 2(c)) but
it deteriorates for longer times (Fig. 2(d)).

In Fig. 3 the evolution of the difference in occupation between the
two sites is presented for $u=\frac{1}{20},N=50$ and $J=1$. We note
also the remarkable agreement between the analytical and numerical
results found for the envelope. The prediction for the rapid oscillations
agrees with the exact results for longer times and more revivals than
in Fig. 2.

\selectlanguage{american}%
\begin{figure}[H]
\selectlanguage{english}%
(a)\qquad{}\qquad{}\qquad{}\qquad{}\qquad{}\qquad{}\qquad{}\qquad{}\qquad{}\qquad{}\qquad{}\qquad{}(b)

\includegraphics[scale=0.45]{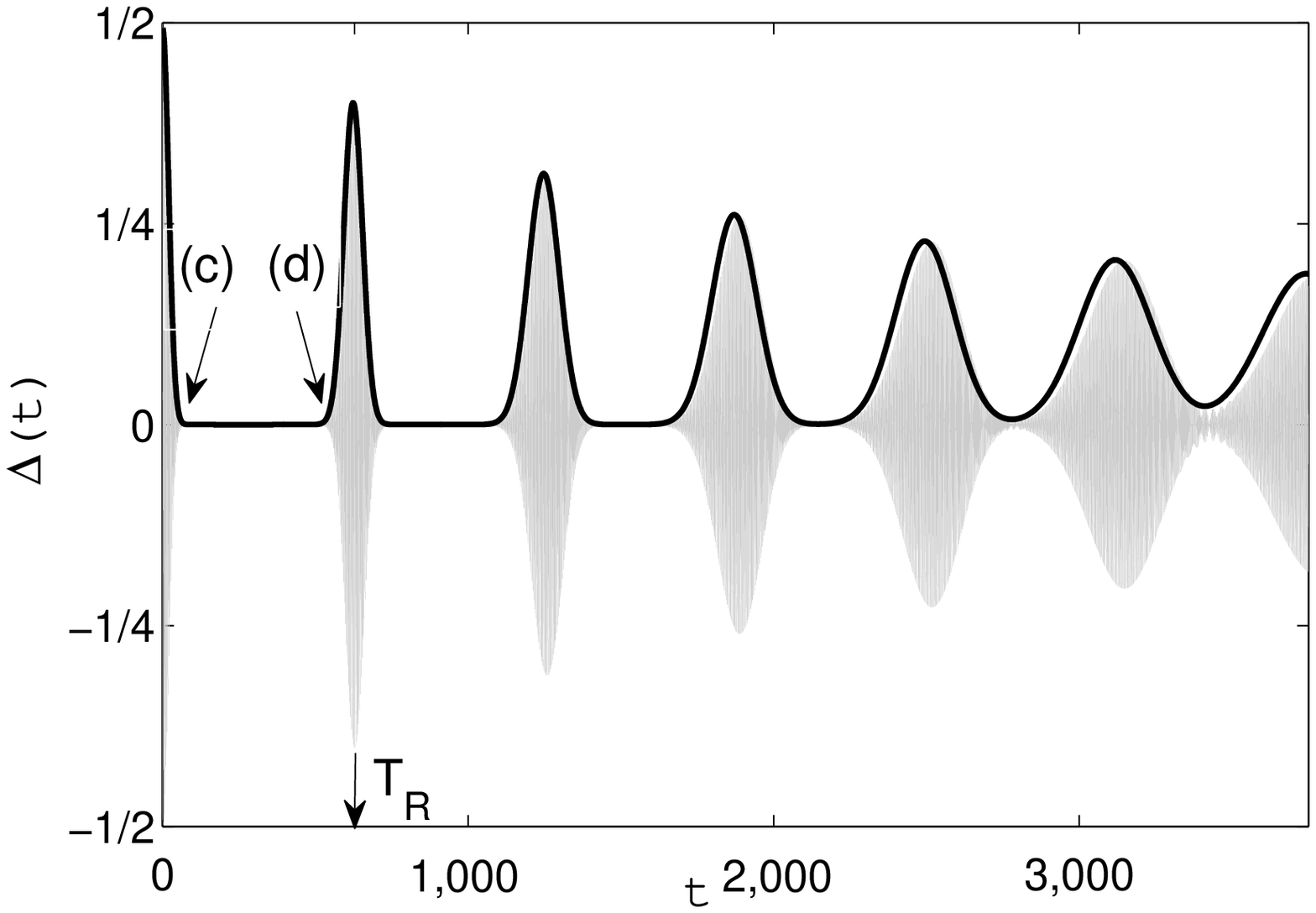}\includegraphics[scale=0.45]{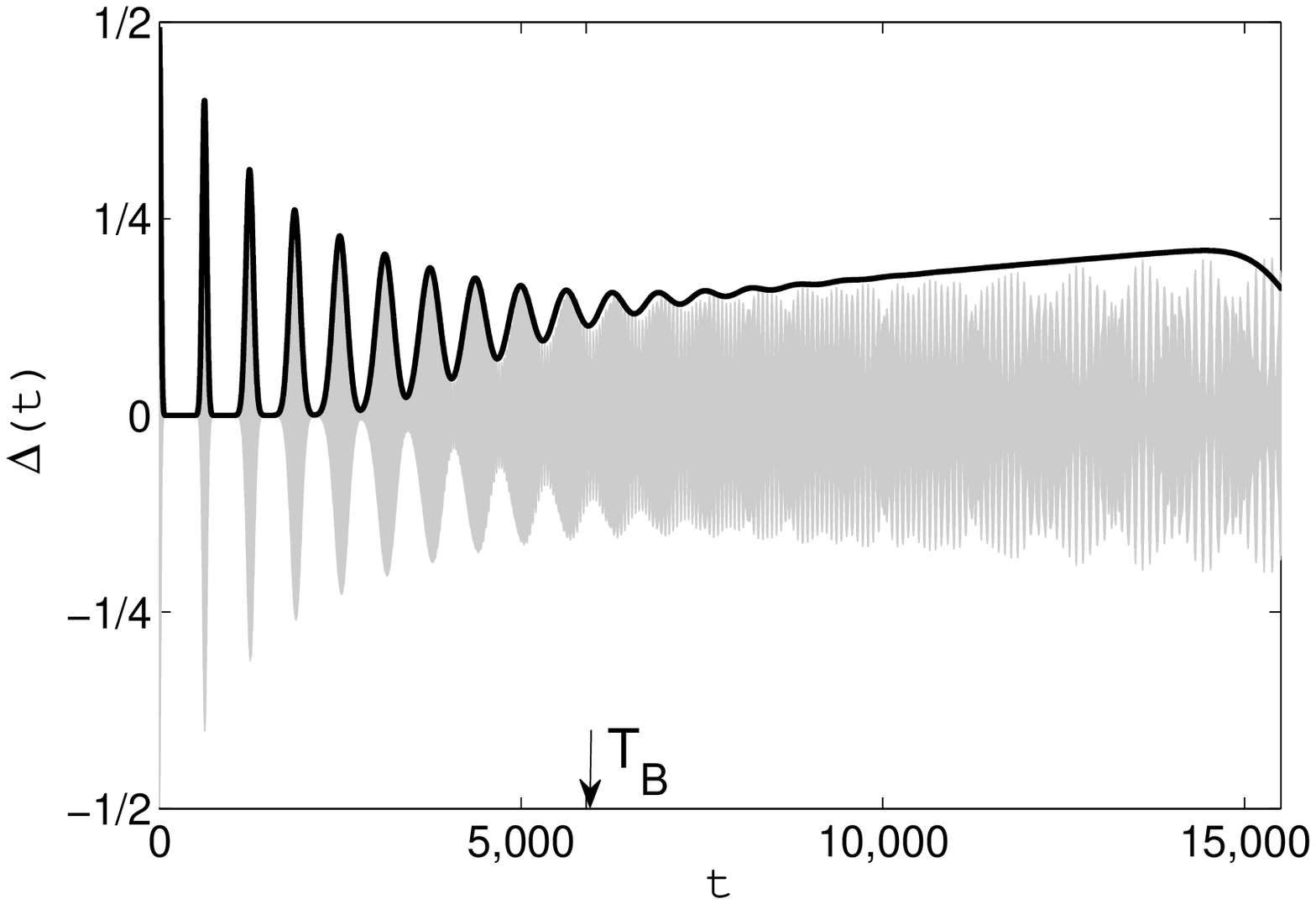}

(c)\qquad{}\qquad{}\qquad{}\qquad{}\qquad{}\qquad{}\qquad{}\qquad{}\qquad{}\qquad{}\qquad{}\qquad{}(d)

\includegraphics[scale=0.45]{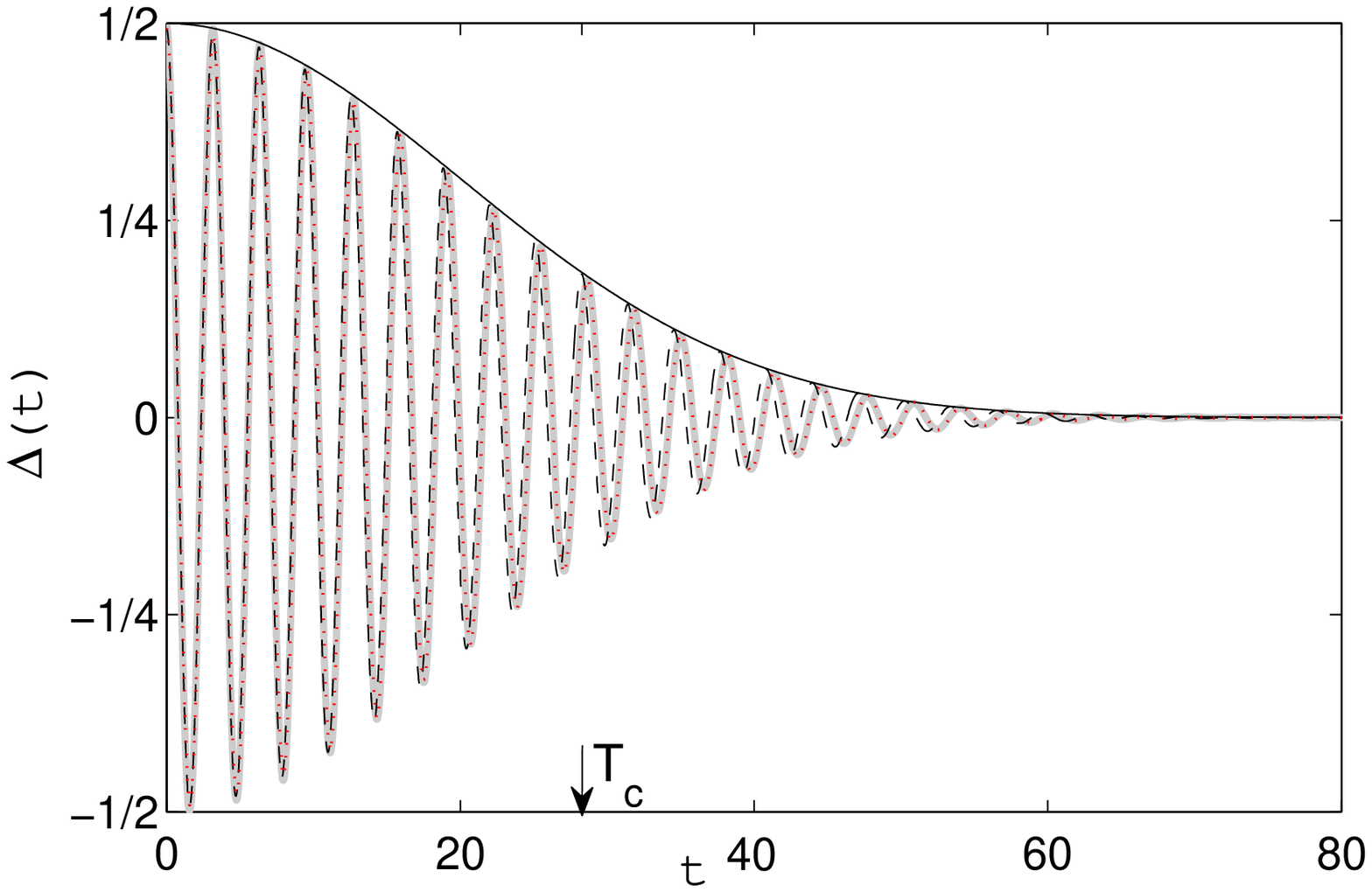}\includegraphics[scale=0.45]{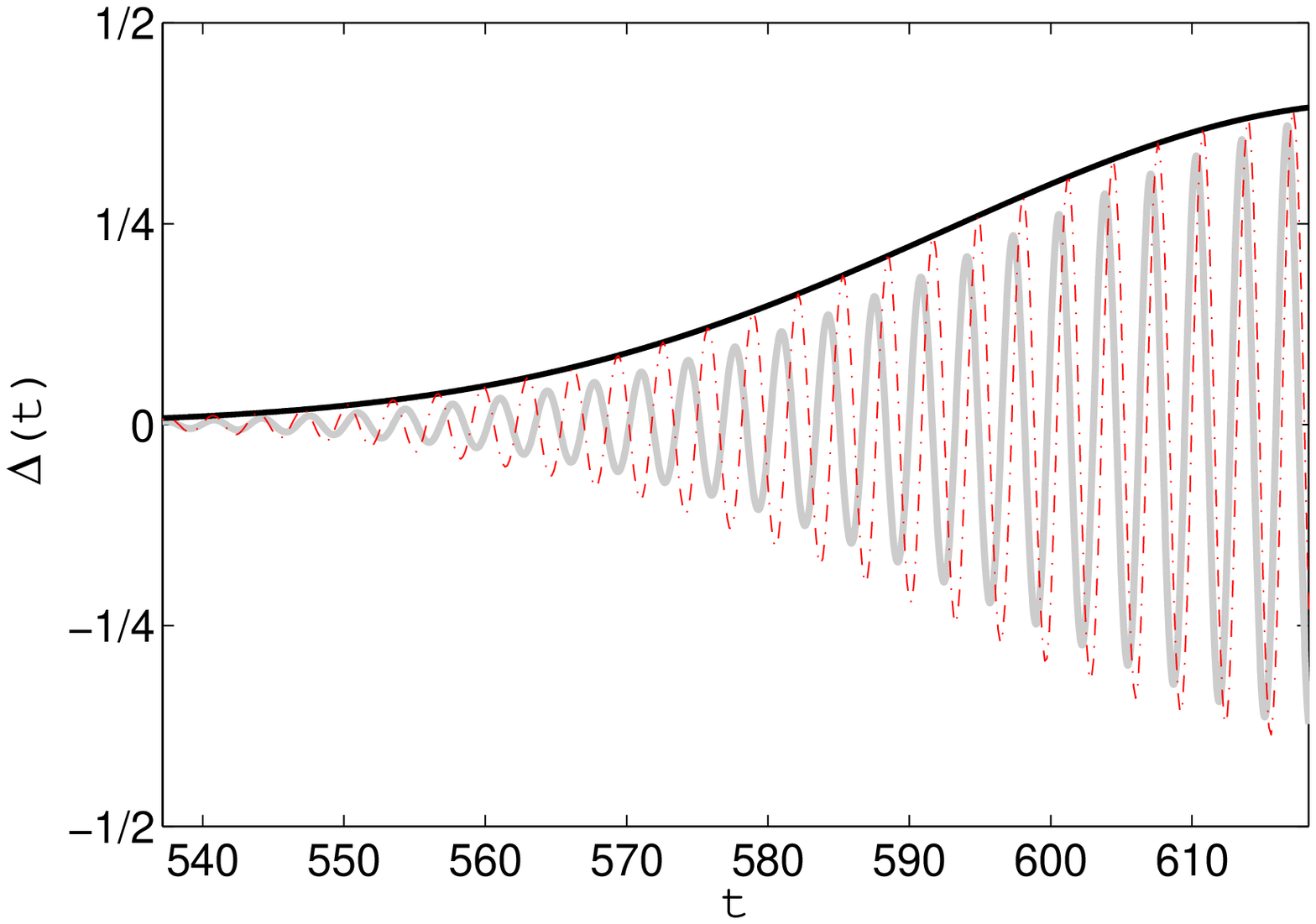}

\selectlanguage{american}%
\caption{\selectlanguage{english}%
(Color online) The normalized difference between the occupation of
the two sites $\Delta\left(t\right)$ for $J=1$, $N=100$ and $u=\frac{1}{2}$.
The light gray line represents the numerical result, obtained by diagonalizing
the Hamiltonian (\ref{eq:H_BH_n}). The black line represents the
envelope based on (\ref{eq:S_z_good}). (a) $\Delta\left(t\right)$
for the time regime $t<T_{B}$. The arrows show the time regimes which
are presented in (c) and (d). The time $T_{R}$ of (\ref{eq:Tr})
is marked. (b) Long time blurring. The time $T_{B}$ where the revivals
mix (see Eqs. (\ref{eq:m_max}) and (\ref{eq:wash_time})) is marked.
(c) Short time dynamics. The red dashed-dot line is given by (\ref{eq:S_z_good})
where $\phi$ and $\phi_{1}$ are given by (\ref{eq:fi_main}) and
(\ref{eq:fi1_main}). The dashed black line presents oscillations
with the unperturbed Rabi's frequency $2J$ (that is approximating
the phase $\phi t-\phi_{1}$ by $2Jt$) and $T_{c}$ of (\ref{eq:Tc})
is marked. (d) the same as (c) for a time interval near the revival
$m=1$, where the analytical result for the phase $\phi_{1}-\phi t$
(\ref{eq:S_z_good}) no longer agrees with the result of exact numerical
calculation.\selectlanguage{american}
}
\end{figure}

\begin{figure}[H]
\selectlanguage{english}%
(a)\qquad{}\qquad{}\qquad{}\qquad{}\qquad{}\qquad{}\qquad{}\qquad{}\qquad{}\qquad{}\qquad{}\qquad{}(b)

\includegraphics[scale=0.45]{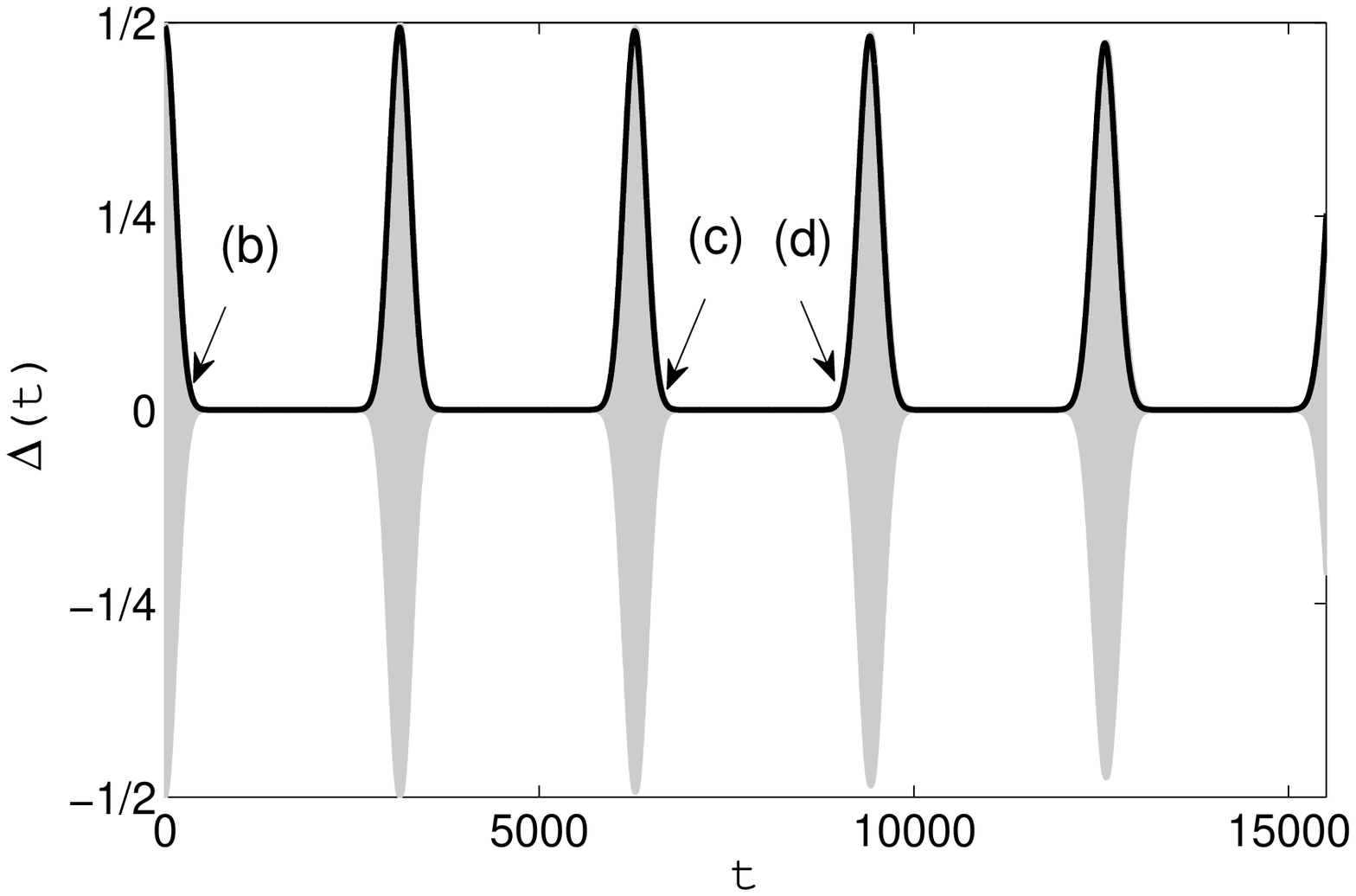}\includegraphics[scale=0.45]{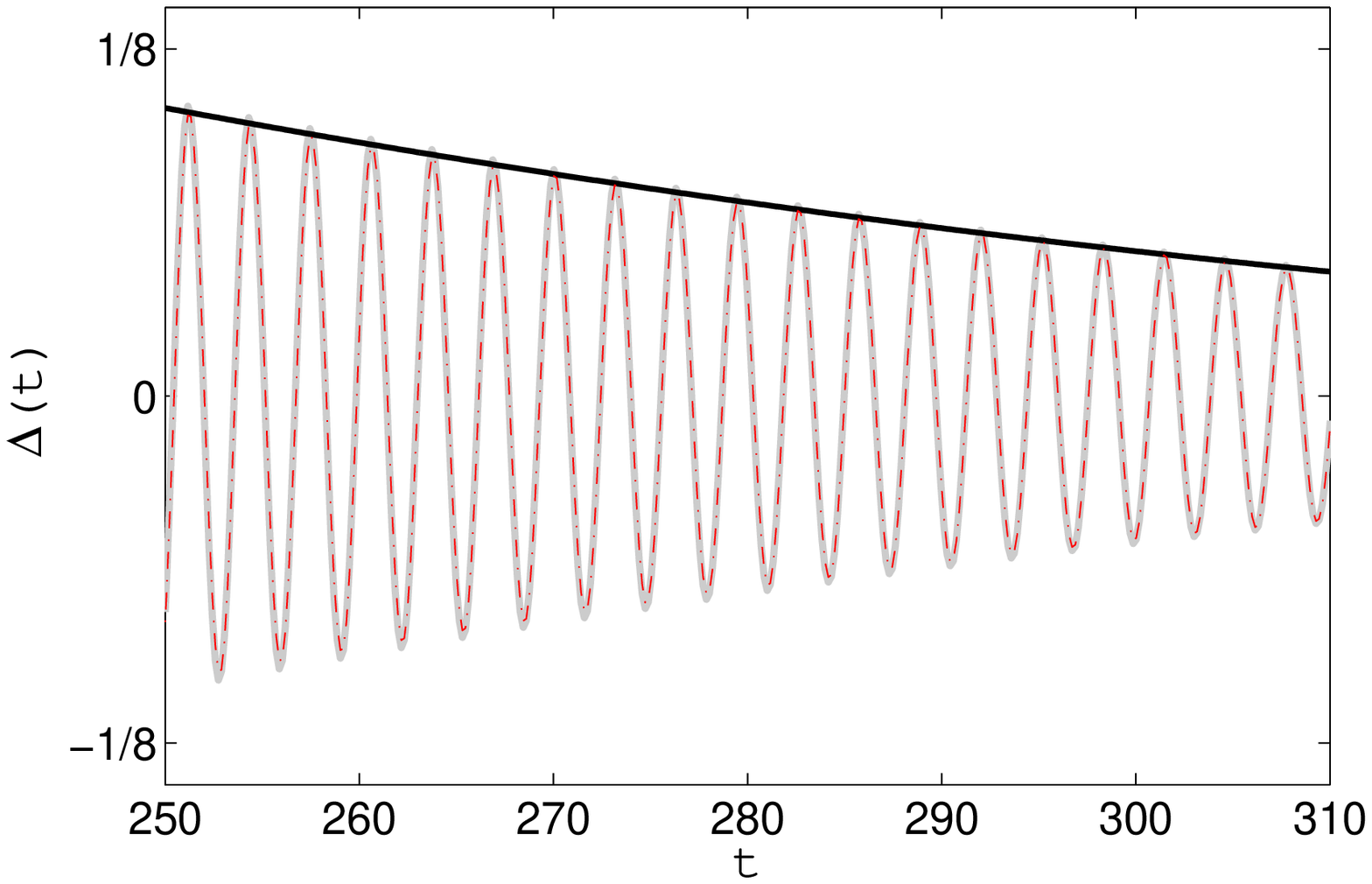}

(c)\qquad{}\qquad{}\qquad{}\qquad{}\qquad{}\qquad{}\qquad{}\qquad{}\qquad{}\qquad{}\qquad{}\qquad{}(d)

\includegraphics[scale=0.45]{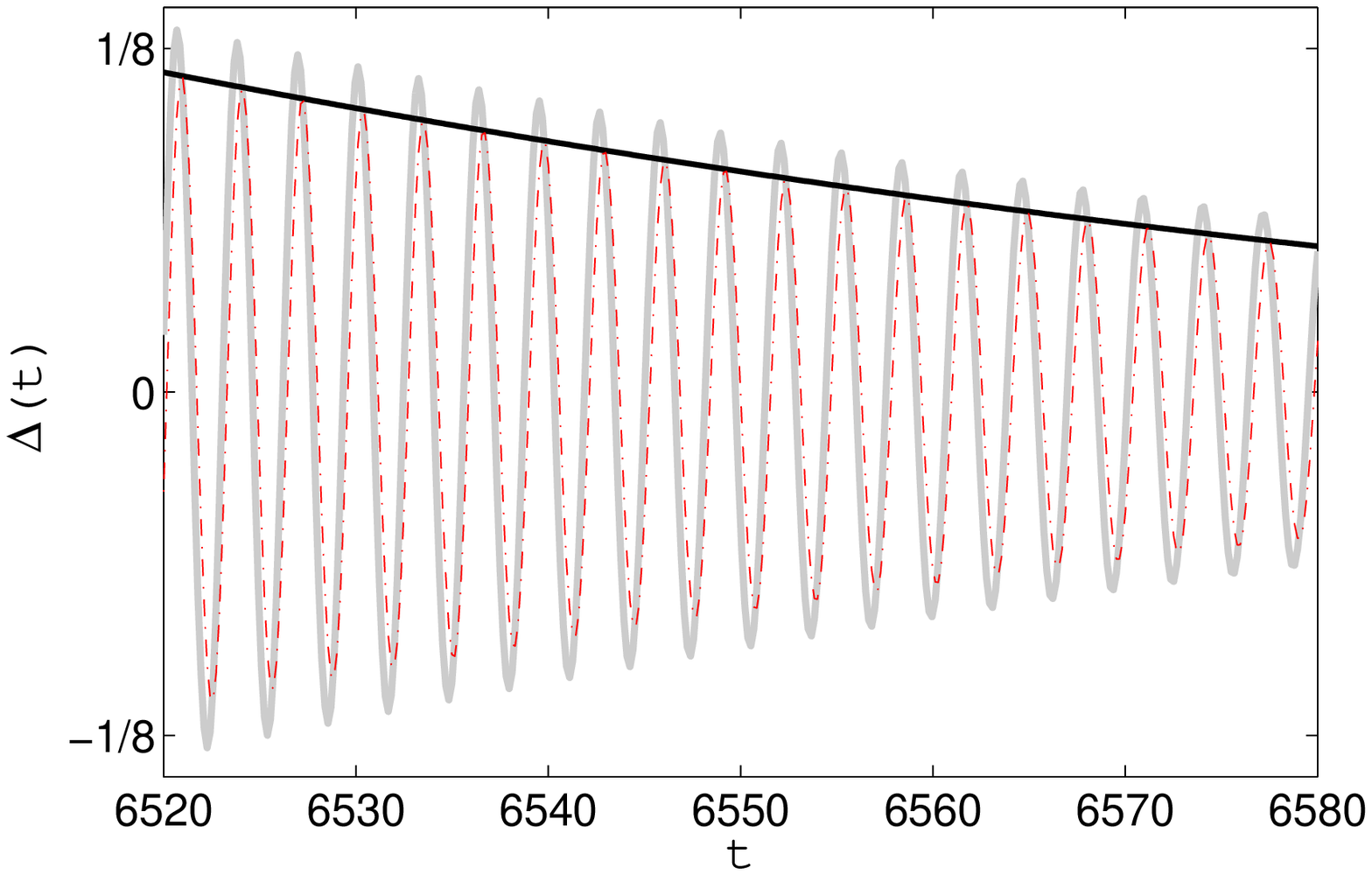}\includegraphics[scale=0.45]{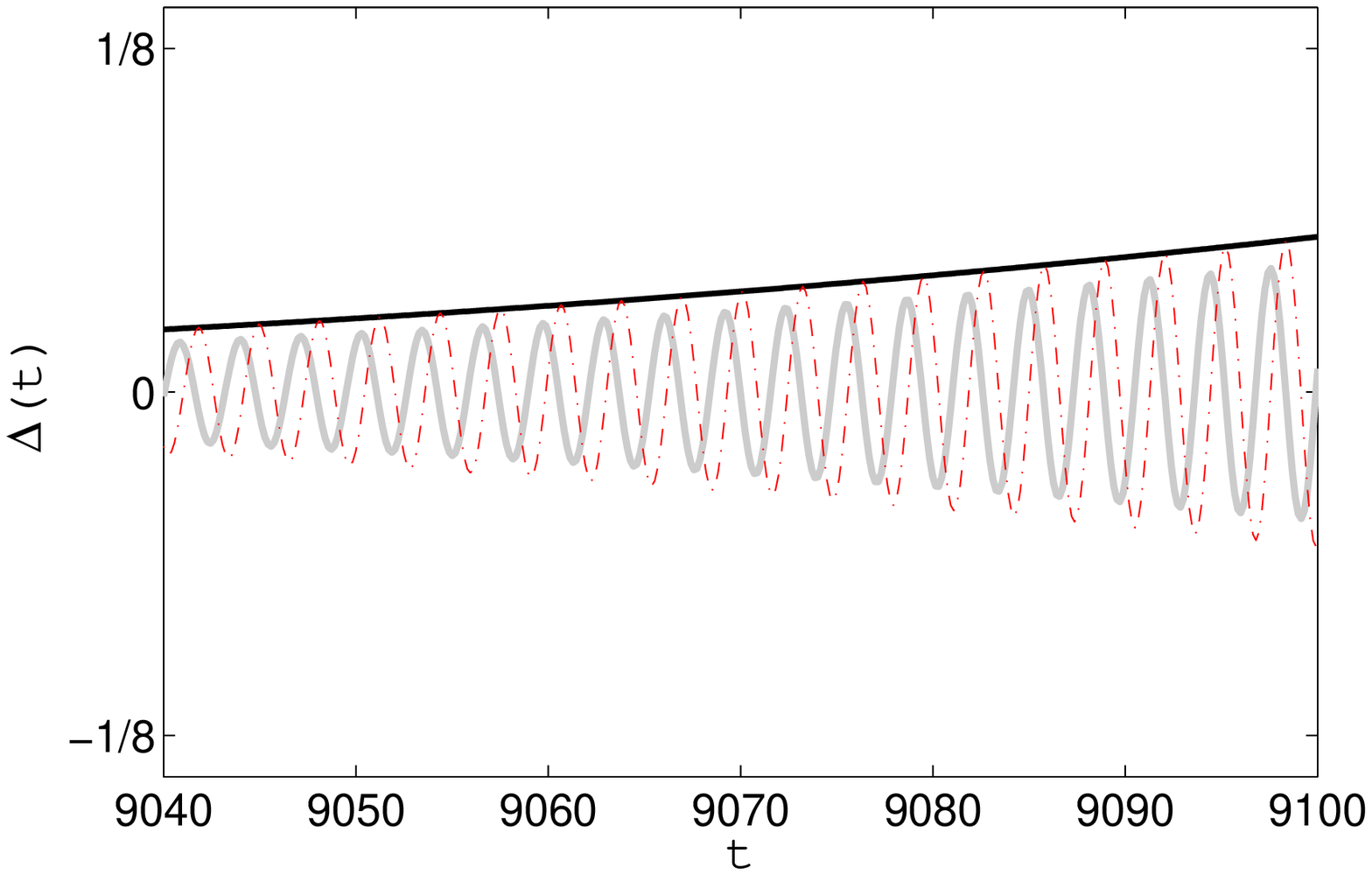}

\selectlanguage{american}%
\caption{\selectlanguage{english}%
(Color online) Similar to Fig. 2 but for $J=1$, $N=50$ and $u=\frac{1}{20}$.
(a)$\Delta\left(t\right)$ for a the time $t<T_{B}$. The arrows show
the time regimes which are presented in (b)-(d). (b) Short time dynamics.
(c) the same as (b) for a time interval near the revival $m=2$. (d)
the same as (b) for a time interval near the revival $m=3$, where
the analytical result for the phase of (\ref{eq:S_z_good}) no longer
agrees with the exact numerical calculation.\selectlanguage{american}
}
\end{figure}
\foreignlanguage{english}{For short times $\left(m=0\right)$, the
dynamics is described by 
\begin{equation}
\Delta\left(t\right)=\left\langle \psi\left|S_{z}\right|\psi\right\rangle =\frac{1}{2}e^{-\frac{1}{2N}J^{2}u^{2}t^{2}}\cos\left(\phi t-\phi_{1}\right)\label{eq:Sz_t}
\end{equation}
and 
\begin{equation}
\left\langle \psi\left(t\right)\left|S_{y}\right|\psi\left(t\right)\right\rangle =-\frac{1}{2}e^{-\frac{1}{2N}J^{2}u^{2}t^{2}}\sin\left(\phi t-\phi_{1}\right).
\end{equation}
Both the expectations of $S_{z}$ and $S_{y}$ oscillate rapidly with
the Rabi frequency $\frac{\pi}{J}$, and at the short time scale have
a Gaussian envelope which is
\begin{equation}
f\left(t\right)=\frac{1}{2}e^{-\frac{1}{2N}J^{2}u^{2}t^{2}}\label{eq:env}
\end{equation}
in the leading order in $u$ and $\frac{1}{N}$. Namely, it decays
on the time scale 
\begin{equation}
T_{c}=\frac{\sqrt{2N}}{Ju}.\label{eq:Tc}
\end{equation}
Note that correction term $J\left(\frac{1}{8}u^{2}+\frac{u}{N}\right)$
to the phase in (\ref{eq:fi_main}) improves the agreement with the
exact numerical results compared to Rabi's phase $2Jt$ (see Fig.
2c). }

For $m>m_{max}$ \foreignlanguage{english}{the} revival peaks overlap
and the picture presented in Figs. 2a and 3a is blurred as demonstrated
in Fig. 2b.

\selectlanguage{english}%

\section{Initial conditions where both sites are occupied}

It is interesting to study the dynamics of a double well where the
initial condition is different occupation of the two wells. Such situation
is encountered, for example, if a condensate is suddenly separated
into two unequal parts, as was done in \cite{jeff}. The initial condition
is of the form
\begin{equation}
\left|\psi\left(t=0\right)\right\rangle =\frac{1}{\sqrt{N!}}\left(a_{R}^{\dagger}\cos\alpha+a_{L}^{\dagger}\sin\alpha\right)^{N}=\frac{1}{2^{N/2}\sqrt{N!}}\left[\left(\cos\alpha+\sin\alpha\right)a_{+}^{\dagger}+\left(\cos\alpha-\sin\alpha\right)a_{-}^{\dagger}\right]^{N}.\label{eq:psi_ns}
\end{equation}
Expansion of (\ref{eq:psi_ns}) as a sum $\sum_{n=0}^{N}c_{n}\left|n\right\rangle $
(with $\left|n\right\rangle $ given by (\ref{eq:S_x_eigen})) yields
\begin{equation}
\begin{array}{ccc}
\left|c_{n}\right| & = & \frac{1}{2^{N/2}}\sqrt{\left(\begin{array}{c}
N\\
\frac{N}{2}+n
\end{array}\right)}\left[\left(\cos\alpha+\sin\alpha\right)^{\frac{N}{2}-n}\left(\cos\alpha-\sin\alpha\right)^{\frac{N}{2}+n}\right].\end{array}
\end{equation}
The coefficients $c_{n}$ of substantial magnitude are distributed
around 
\begin{equation}
n_{max}=\frac{N}{2}\sin\left(2\alpha\right)\label{eq:n_max}
\end{equation}
so that
\begin{equation}
c_{n}=\left(\frac{2\beta}{\pi N}\right)^{\frac{1}{4}}e^{-\frac{1}{N}\beta\left(n-n_{max}\right)^{2}}\label{eq:cna}
\end{equation}
 and
\begin{equation}
c_{n}c_{n+1}=\sqrt{\frac{2\beta}{\pi N}}e^{-\frac{2}{N}\beta\left[\left(n-n_{max}\right)^{2}+\left(n-n_{max}\right)+\frac{1}{2}\right]}.\label{eq:CnCn1}
\end{equation}
where 
\begin{equation}
\beta=\frac{1}{\cos^{2}\left(2\alpha\right)}.\label{eq:betta}
\end{equation}
The expectation value $\Delta\left(t\right)=\left\langle S_{z}\left(t\right)\right\rangle $
is calculated in a similar way to what was done in the previous section.
The differences are:
\begin{enumerate}
\item $\frac{n_{max}}{N}$ is not necessarily negligible and therefore $\widetilde{S}_{+}\left|n\right\rangle =\sqrt{\left(\frac{N}{2}+n+1\right)\left(\frac{N}{2}-n\right)}\left|n+1\right\rangle \approx\sqrt{\frac{N^{2}}{4}-n_{max}^{2}}\left|n+1\right\rangle $
and not $\frac{N}{2}\left|n+1\right\rangle $ (see for comparison
(\ref{eq:s+})) .
\item Due to (\ref{eq:CnCn1}), $\Delta\left(t\right)=\left\langle S_{z}\left(t\right)\right\rangle $
is multiplied by $\sqrt{\beta}$.
\item The $\beta$ in the exponent of (\ref{eq:CnCn1}) affects the result
of the integral $\widetilde{S}_{m}$ of (\ref{eq:Sm_int}), see App.
B.
\item For $n\approx n_{max}$, it is possible that $\frac{3}{2N^{2}}u^{2}n^{2}$
in (\ref{eq:Spect_diff2}) is not negligible compared to $\frac{2}{N}un$.
Consequently, the revival time $T_{R}$ will be modified as described
in what follows. We substitute in (\ref{eq:Spect_diff2}) $n=n_{max}+\Delta n$
and write $E_{n}^{\left(BH2\right)}-E_{n+1}^{\left(BH2\right)}=J\left(2+\frac{2}{N}u\left(n_{max}+\Delta n\right)+\frac{1}{8}u^{2}-\frac{3}{2N^{2}}u^{2}\left(n_{max}+\Delta n\right)^{2}\right)$
for $n_{max}\gg1$. The first constructive interference is obtained
for $J\left(\frac{2}{N}u\Delta n-\frac{3}{N^{2}}u^{2}n_{max}\Delta n\right)T_{R}=2\pi$,
namely 
\begin{equation}
T_{R}=\gamma\frac{\pi N}{uJ}\label{eq:Trg}
\end{equation}
 where $\gamma=\left(1-\frac{3}{2}u\frac{n_{max}}{N}\right)^{-1}=\left(1-\frac{3}{4}u\sin\left(2\alpha\right)\right)^{-1}$.
\end{enumerate}
Therefore, for the initial condition (\ref{eq:psi_ns}), the expectation
value $\Delta\left(t\right)$ takes the form (as can be seen by modifying
(\ref{eq:S_z_good})),
\begin{eqnarray}
\Delta\left(t\right) & = & \frac{\sqrt{\beta\left(\frac{N^{2}}{4}-n_{max}^{2}\right)}}{2}\cdot\label{eq:S_z_good-1}\\
 &  & \sum_{m}\frac{e^{-\frac{u^{2}}{32}}}{\left[\beta^{2}+\frac{9}{16}u^{2}m^{2}\gamma^{2}\pi^{2}\right]^{1/4}}\exp\left[\frac{-\frac{1}{2}J^{2}u^{2}\beta\left(t+\frac{3m\pi}{2J}-mT_{R}\right)^{2}+\bar{\eta}}{N\left(\beta^{2}+\frac{9}{16}u^{2}m^{2}\gamma^{2}\pi^{2}\right)}\right]\cos\left(\bar{\phi}_{1}-\phi t\right)
\end{eqnarray}
where
\begin{equation}
\bar{\eta}=\frac{\beta^{3}}{2}+\frac{9}{8}\beta u^{2}m^{2}\pi^{2},\label{eq:L-1}
\end{equation}
\begin{equation}
\bar{\phi}_{1}\approx\frac{u^{2}}{8\beta}\left(2J\tau+\frac{3}{2}m\gamma\pi\right)+u\left(\frac{J\tau\beta^{2}}{N}+\frac{3}{8\beta}\left(m\gamma\pi+\frac{J}{N}u\tau\right)\right)\label{eq:fi1_main-1}
\end{equation}
and $\phi$ is given by (\ref{eq:fi_main}). 

We turn now to estimate the conditions for the validity of the approximation
(\ref{eq:S_z_good-1}). The width of the Gaussian (\ref{eq:CnCn1})
is $\sqrt{\frac{N}{\beta}},$ therefore it is required that
\begin{equation}
\frac{N}{2}-n_{max}>\sqrt{\frac{N}{\beta}},
\end{equation}
therefore by (\ref{eq:n_max}),
\begin{equation}
\varepsilon_{1}\equiv\frac{1}{\sqrt{N}}\left(\frac{2}{1-\sin\left(2\alpha\right)}\right)\left|\cos\left(2\alpha\right)\right|<1.\label{eq:N>}
\end{equation}
The result (\ref{eq:N>}) is demonstrated in Fig. 4.

In addition, the perturbation theory in $u$ adds to $\widetilde{S}_{m}$
a term of the form $\left(\frac{un_{max}}{N}\right)\widetilde{S}_{m}$
(see App. C (\ref{eq:S1-1}), where the first order correction is
calculated). Therefore, the higher orders can be neglected only if
$un_{max}\ll N$, namely (see (\ref{eq:n_max})), 
\begin{equation}
\varepsilon_{2}\equiv\frac{un_{max}}{N}=\frac{u}{2}\sin\left(2\alpha\right)\ll1.\label{eq:N<}
\end{equation}
Furthermore, the spectrum (\ref{eq:E_BH2}) is more accurate for small
values of $\left|n\right|$ (see Fig. 1) where $H$ in (\ref{eq:Sx})
is small. If one wants to describe the dynamics for $\varepsilon_{2}\lesssim1$,
higher orders in the expansion of (\ref{eq:Sx}) might be needed.

\selectlanguage{american}%
\begin{figure}[H]
\selectlanguage{english}%
(a)\qquad{}\qquad{}\qquad{}\qquad{}\qquad{}\qquad{}\qquad{}\qquad{}\qquad{}\qquad{}\qquad{}\qquad{}(b)

\includegraphics[scale=0.45]{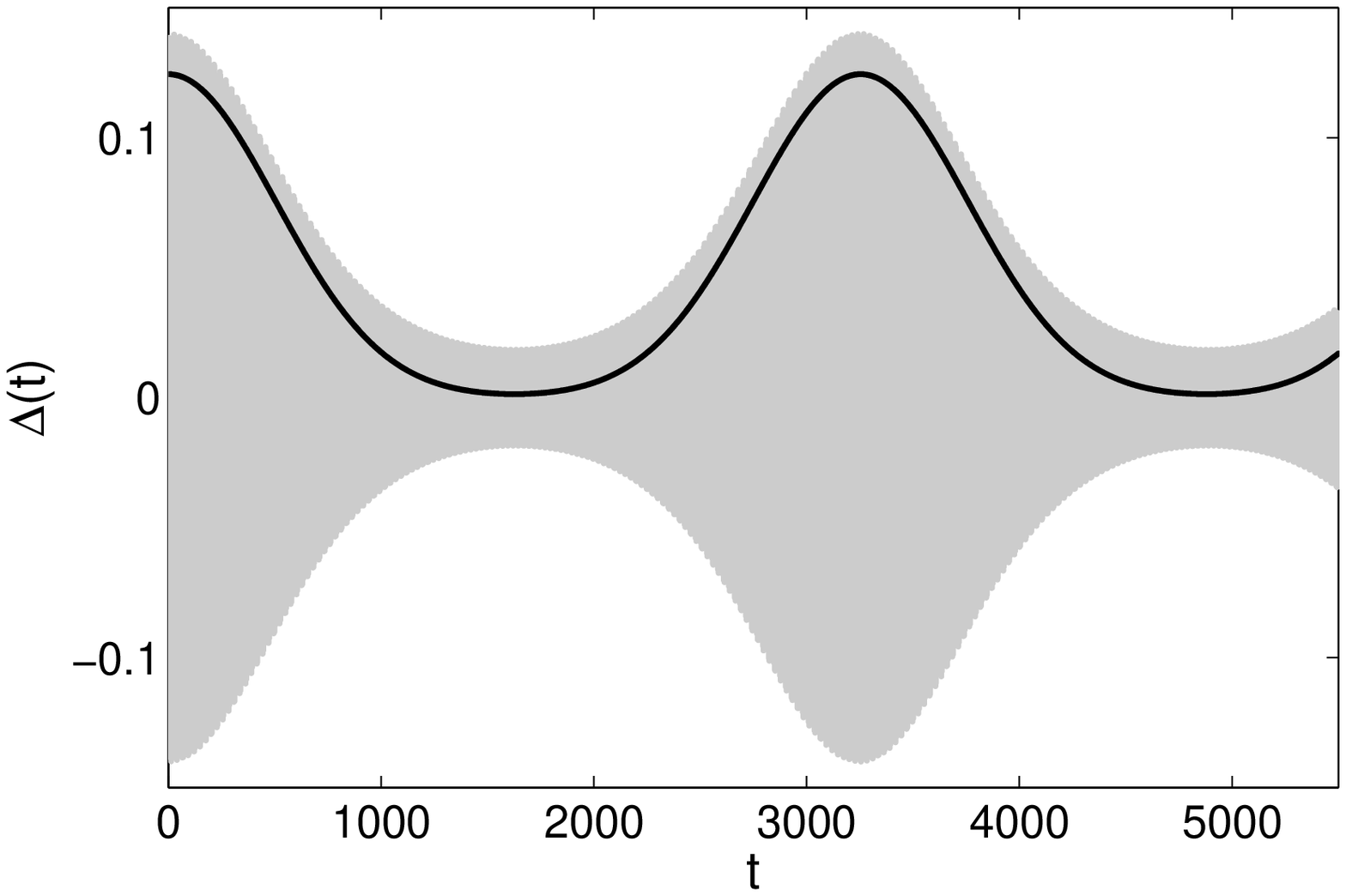}\includegraphics[scale=0.45]{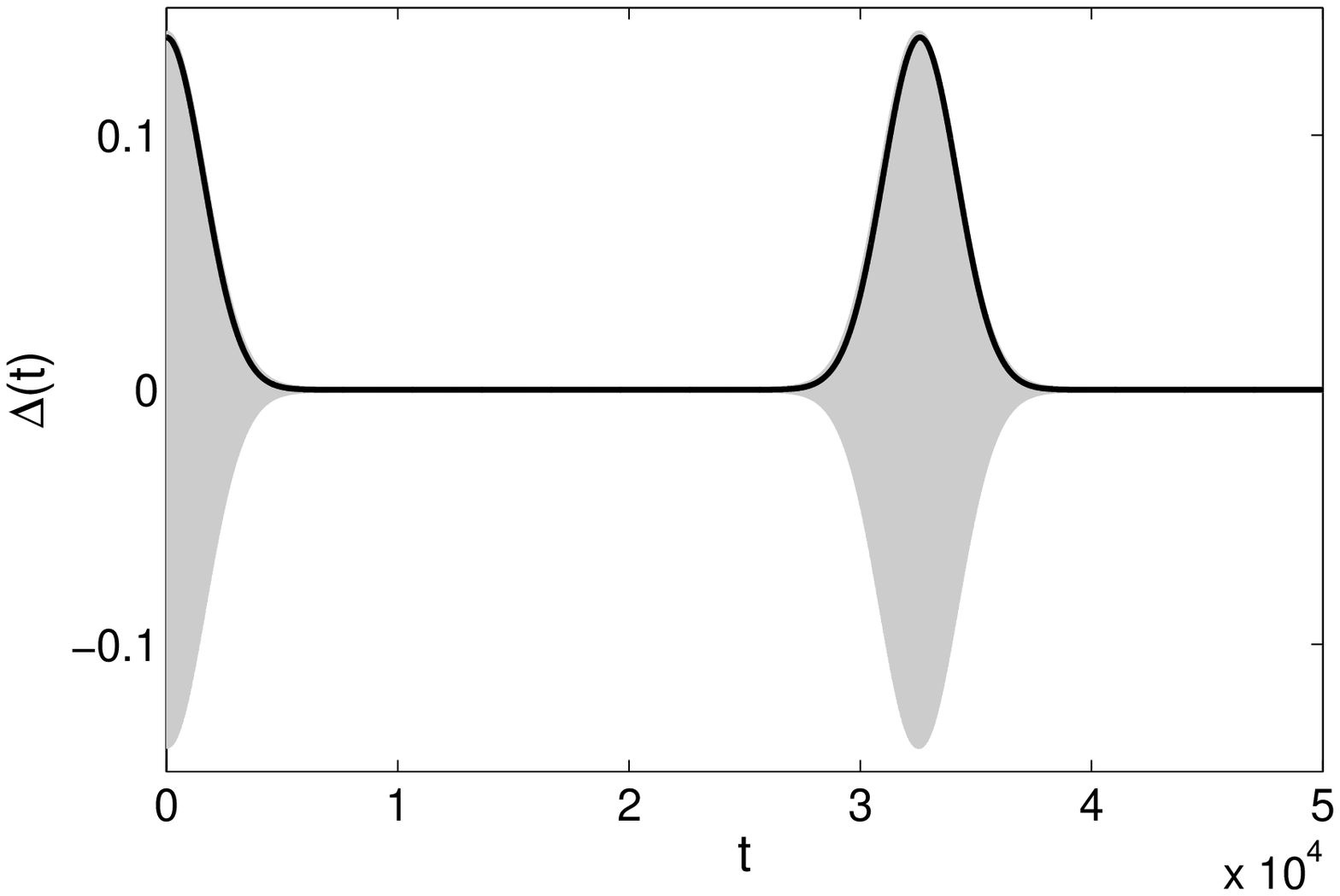}

\selectlanguage{american}%
\caption{\selectlanguage{english}%
The normalized difference between the occupation of the two sites
$\Delta\left(t\right)$ for the initial condition (\ref{eq:psi_ns})
with $\cos\alpha=\frac{3}{5}$, $\gamma=1.0373$. The parameters are
$J=1$ and $u=\frac{1}{20}$, namely $\varepsilon_{2}=0.024$. The
light gray line represents the exact numerical result, obtained by
diagonalizing the Hamiltonian (\ref{eq:H_BH}) and the black line
represents the envelope based on (\ref{eq:S_z_good-1}). The numbers
of particles are (a) $N=50$ (where $\varepsilon_{1}=14$) (b) $N=500$
(where $\varepsilon_{1}=0.63$).\selectlanguage{american}
}
\end{figure}

\selectlanguage{english}%

\section{Summary and discussion}

In the present work the dynamics of the two site Bose Hubbard model
defined by (\ref{eq:H_BH_n}) and (\ref{eq:H_BH}) were analyzed.
We analyzed it for weak coupling $u$ (\ref{eq:u}) and for large
number of particles $N$. The calculation was preformed to order $u^{2}$
and to the leading order in $\frac{1}{N}$, using a semiclassical
method where $\frac{1}{N}$ plays the role of the Planck's constant.
It is important to note that this is not the standard quantum perturbation
theory that requires $uN<O\left(1\right)$ but here it is requires
only that $u\leq O\left(1\right)$.

In particular, the normalized difference in the occupation of the
sites $\Delta\left(t\right)=\left\langle \psi\left(t\right)\left|S_{z}\right|\psi\left(t\right)\right\rangle $
as a function of time was calculated in a situation where initially
all bosons are on one site leading to (\ref{eq:S_z_good}) with (\ref{eq:L}),
(\ref{eq:fi_main}) and (\ref{eq:fi1_main}). It is compared to the
exact numerical solution in Figs. 2 and 3. For the envelope, remarkable
agreement with the exact numerical solution is found. The solution
exhibits rapid (Rabi) oscillations. The quality of the analytical
result for these oscillations is initially very good but it deteriorates
with time. The normalized population difference exhibits three time
scales: $T_{c}$ (\ref{eq:Tc}), $T_{R}$ (\ref{eq:Tr}) and $T_{B}$
(\ref{eq:wash_time}). Initially, it collapses at a time $T_{c}$
given by (\ref{eq:Tc}). Then, it exhibits revivals at times $mT_{R}$
with $T_{R}$ given by (\ref{eq:Tr}). These revivals are of increasing
width (\ref{eq:dtRm}). Eventually, at $T_{B}$ given by (\ref{eq:wash_time}),
this picture is washed away.

Comparison between the approximate result and the exact numerical
calculation demonstrates that the result obtained indeed requires
the terms in order $u^{2}$ and $\frac{1}{N}$. The classical approximation
(\ref{eq:H_sx_fi}) reproduces correctly the rapid oscillations for
short times. Such a behavior is found also for the GPE in double well
\cite{Milburn,Smerzi}. Quantization is essential for the collapses
and revivals. The collapse and revival times are predicted correctly
by the first order in the interaction $u$, however for the width
of the peaks the order $u^{2}$ is required, since the width depends
on $m$ via the combination $m^{2}u^{2}$.

Collapses and revivals were found in various situations \cite{Milburn,Bloch_colapse,Lewenstein,Wright,wright&walls,castin&dalibard,Mark,Talbot,Berry&Sclich,carpet,Rayleigh,Eberly}.
To the best of our knowledge, (\ref{eq:S_z_good}) is the only complete
analytic description of this situation for a specific model of interacting
bosons. It is reminiscent of the dynamics of the Jaynes-Cummings model
\cite{Eberly}.

We studied also the case where initially both sites are populated
and found out an approximation that is good if the initial difference
in occupation is sufficiently large. For finer details see Eqs. (\ref{eq:N>}),
(\ref{eq:N<}) and Fig. 4. In this case, collapses and revivals are
found as well where also here the collapse time $T_{c}$ and the width
of the reviving peaks are proportional to $\sqrt{N}$ and the revival
time is proportional to $N$. However, the revival time depends on
the initial condition, as is seen from (\ref{eq:Trg}).

The generalization to other situations is left for further work.
\begin{acknowledgments}
This work resulted of a discussion with D. Cohen on ref. \cite{Doron}.
We thank him for motivating this direction of research and many critical
discussions and communications. We thank also O. Alon, O. Alus, I.
Bloch, E. Shimshoni and J. Steinhauer for illuminating and informative
discussions. The work was supported in part \foreignlanguage{american}{by
the Israel Science Foundation (ISF) grant number 1028/12, by the US-Israel
Binational Science Foundation (BSF) grant number 2010132 and by the
Shlomo Kaplansky academic chair.}
\end{acknowledgments}

\section*{Appendix A}

In this appendix, we relate the BH Hamiltonian (\ref{eq:H_BH}) to
the spin Hamiltonian (\ref{eq:H_s}). Substituting the definitions
(\ref{eq:S_def}) in (\ref{eq:H_s}), we get
\begin{equation}
H'=-J\left(a_{R}^{\dagger}a_{L}+a_{R}^{\dagger}a_{L}\right)+\frac{U}{2}\left(a_{L}^{\dagger}a_{L}-a_{R}^{\dagger}a_{R}\right)^{2}.
\end{equation}
The first term of $H'$ is identical to the first term of $H_{BH}$.
The second term is
\begin{equation}
\begin{array}{ccc}
\frac{U}{2}\left(a_{L}^{\dagger}a_{L}-a_{R}^{\dagger}a_{R}\right)^{2} & = & \frac{U}{2}\left(a_{R}^{\dagger}a_{R}a_{R}^{\dagger}a_{R}-2a_{L}^{\dagger}a_{L}a_{R}^{\dagger}a_{R}+a_{L}^{\dagger}a_{L}a_{L}^{\dagger}a_{L}\right)\\
 & = & \frac{U}{2}\left(a_{R}^{\dagger}a_{R}^{\dagger}a_{R}a_{R}+a_{R}^{\dagger}a_{R}-2a_{L}^{\dagger}a_{L}a_{R}^{\dagger}a_{R}+a_{L}^{\dagger}a_{L}^{\dagger}a_{L}a_{L}+a_{L}^{\dagger}a_{L}\right).
\end{array}\label{eq:U_term}
\end{equation}
We find that
\begin{equation}
\begin{array}{ccl}
H_{BH}-H' & = & \frac{U}{2}\left(a_{R}^{\dagger}a_{R}^{\dagger}a_{R}a_{R}-a_{R}^{\dagger}a_{R}+2a_{L}^{\dagger}a_{L}a_{R}^{\dagger}a_{R}+a_{L}^{\dagger}a_{L}^{\dagger}a_{L}a_{L}-a_{L}^{\dagger}a_{L}\right)\\
 & = & \frac{U}{2}\left(a_{R}^{\dagger}a_{R}a_{R}^{\dagger}a_{R}+2a_{L}^{\dagger}a_{L}a_{R}^{\dagger}a_{R}+a_{L}^{\dagger}a_{L}a_{L}^{\dagger}a_{L}-2a_{R}^{\dagger}a_{R}-2a_{L}^{\dagger}a_{L}\right)\\
 & = & \frac{1}{2}N^{2}U-NU.
\end{array}
\end{equation}
Therefore,
\begin{equation}
\begin{array}{ccc}
H_{BH} & = & H'+\frac{1}{2}JuN-Ju\end{array}.\label{eq:constants}
\end{equation}
That reduces to (\ref{eq:H_s}) up to a constant.

\section*{Appendix B}

In this appendix, we calculate the integral (\ref{eq:Sm_int}) and
the corresponding integral required in Sec. VI, which are of the form
\begin{equation}
\widetilde{S}'{}_{m}=\int_{-\infty}^{\infty}e^{-\left(Ax^{2}+Bx\right)}dx=\sqrt{\frac{\pi}{A}}e^{B^{2}/4A}\label{eq:g_int}
\end{equation}
where
\begin{equation}
\widetilde{S}{}_{m}=\frac{\sqrt{2}e^{-\frac{\beta}{N}}}{\sqrt{\pi N}}\widetilde{S}'{}_{m}
\end{equation}
\begin{equation}
\begin{array}{ccl}
A & =\frac{1}{N} & \left[2\beta-\frac{3}{2}iu\left(\bar{m}\pi+\frac{J}{N}u\tau\right)\right]\\
B & = & \frac{2}{N}\left[\beta+iJu\tau\right]\\
\bar{m} & = & \gamma m.
\end{array}\label{eq:AB}
\end{equation}
In Sec. V we consider the case $\beta=1,\gamma=1$ while in section
VI, $\beta=\frac{1}{\cos^{2}\left(2\alpha\right)}$ and $\gamma=\left[1-\frac{3}{4}u\sin\left(2\alpha\right)\right]^{-1}$.
There, $\alpha$ determines the initial conditions, see (\ref{eq:psi_ns}).
In order to write (\ref{eq:g_int}) explicitly, we preform some manipulations
where for each order of $\tau$, only the dominant order in $u$ is
taken into account.
\begin{equation}
\frac{B^{2}}{4A}=\frac{\frac{1}{N}\left[\beta^{2}+2iJu\tau\beta-J^{2}u^{2}\tau^{2}\right]}{2\beta-\frac{3}{2}iu\left(\bar{m}\pi+\frac{J}{N}u\tau\right)}.
\end{equation}
After multiplying the numerator and the denominator by the complex
conjugate of the denominator,
\begin{eqnarray*}
\frac{B^{2}}{4A} & = & \frac{\left[\beta^{2}+2iJu\tau\beta-J^{2}u^{2}\tau^{2}\right]\left[2\beta+\frac{3}{2}iu\left(\bar{m}\pi+\frac{J}{N}u\tau\right)\right]}{N\left[4\beta^{2}+\frac{9}{4}u^{2}\left(\bar{m}\pi+\frac{J}{N}u\tau\right)^{2}\right]}\\
 & = & \frac{2\beta^{3}+4iJu\tau\beta^{2}-2J^{2}u^{2}\tau^{2}\beta+\frac{3}{2}iu\beta^{2}\left(\bar{m}\pi+\frac{J}{N}u\tau\right)}{N\left[4\beta^{2}+\frac{9}{4}u^{2}\left(\bar{m}\pi+\frac{J}{N}u\tau\right)^{2}\right]}\\
 &  & -\frac{3Ju^{2}\tau\beta\left(\bar{m}\pi+\frac{J}{N}u\tau\right)+\frac{3}{2}iJ^{2}u^{3}\tau^{2}\left(\bar{m}\pi+\frac{J}{N}u\tau\right)}{N\left[4\beta^{2}+\frac{9}{4}u^{2}\left(\bar{m}\pi+\frac{J}{N}u\tau\right)^{2}\right]}\\
 & = & \frac{2\beta^{3}-3Ju^{2}\tau\bar{m}\pi\beta-\beta\left(2J^{2}u^{2}+\frac{3}{N}J^{2}u^{3}\right)\tau^{2}}{N\left[4\beta^{2}+\frac{9}{4}u^{2}\left(\bar{m}\pi+\frac{J}{N}u\tau\right)^{2}\right]}\\
 &  & +\frac{i\left[4Ju\tau\beta^{2}+\frac{3}{2}u\beta^{2}\left(\bar{m}\pi+\frac{J}{N}u\tau\right)-\frac{3}{2}Ju^{3}\tau^{2}\left(\bar{m}\pi+\frac{J}{N}u\tau\right)\right]}{N\left[4\beta^{2}+\frac{9}{4}u^{2}\left(\bar{m}\pi+\frac{J}{N}u\tau\right)^{2}\right]}.
\end{eqnarray*}
To the leading order in $u$,
\[
\frac{B^{2}}{4A}=\frac{D_{R}+iD_{I}}{D_{D}}
\]
where
\begin{equation}
D_{R}=\frac{1}{N}\left\{ 2\beta^{3}-3Ju^{2}\tau\bar{m}\pi\beta-2\beta J^{2}u^{2}\tau^{2}\right\} \label{eq:DR}
\end{equation}
\begin{equation}
D_{I}=\frac{1}{N}\left[4Ju\tau\beta^{2}+\frac{3}{2}u\beta^{2}\left(\bar{m}\pi+\frac{J}{N}u\tau\right)-\frac{3}{2}J^{2}u^{3}\tau^{2}\left(\bar{m}\pi+\frac{J}{N}u\tau\right)\right]
\end{equation}
\begin{equation}
\begin{array}{ccl}
D_{D} & = & 4\beta^{2}+\frac{9}{4}u^{2}\left(\bar{m}\pi+\frac{J}{N}u\tau\right)^{2}\end{array}.\label{eq:DDD}
\end{equation}
$D_{R}$ can be written as
\begin{equation}
D_{R}=-\frac{2\beta}{N}J^{2}u^{2}\left(\tau+\frac{3\bar{m}\pi}{2J}\right)^{2}+\frac{2\beta^{3}}{N}+\frac{9}{2N}u^{2}\bar{m}^{2}\pi^{2}\beta.
\end{equation}
Now we turn to calculate $\sqrt{\frac{\pi}{A}}$ which appears in
(\ref{eq:g_int}). According to (\ref{eq:DDD}), 
\begin{equation}
A=\frac{\sqrt{D_{D}}}{N}e^{i\phi_{A}}
\end{equation}
where
\begin{equation}
\tan\phi_{A}=-\frac{3}{4\beta}u\left(\bar{m}\cdot\pi+\frac{J}{N}u\tau\right).
\end{equation}
Therefore,
\begin{equation}
\sqrt{\frac{\pi}{A}}=\frac{\sqrt{\pi N}}{D_{D}^{1/4}}e^{-i\frac{\phi_{A}}{2}}
\end{equation}
and
\begin{equation}
\widetilde{S}_{m}=\frac{\sqrt{2}}{D_{D}^{1/4}}e^{\frac{D_{R}}{D_{D}}-\frac{\beta}{N}+i\phi_{S}}
\end{equation}
where
\begin{equation}
\phi_{s}=\frac{D_{I}}{D_{D}}-\frac{\phi_{A}}{2}.\label{eq:fS}
\end{equation}
To the order $u^{2}$,
\begin{equation}
\phi_{A}\approx-\frac{3}{4\beta}u\left(\bar{m}\cdot\pi+\frac{J}{N}u\tau\right)
\end{equation}
and
\begin{equation}
\phi_{s}=u\left(\frac{J\tau}{N}+\frac{3}{8}\left(\frac{1}{\beta}+\frac{1}{N}\right)\left(\bar{m}\cdot\pi+\frac{J}{N}u\tau\right)\right).
\end{equation}

\section*{Appendix C}

In this appendix we calculate the correction resulting from the fact
that for $u\neq0$ the eigenstates of $H=-S_{x}+uS_{z}^{2}$ are not
identical to the eigenstates of $S_{x}$. Perturbation theory is justified
only for $uN<1$ because the typical spacing between eigenvalues of
$S_{x}$ is about $\frac{1}{N}$ while the maximum of the perturbation
$uS_{z}^{2}$ is about $\frac{u}{4}$. However, most of the results
presented in Sec. V-VI are applicable for $u<1$ (where it is possible
that $uN\gg1$). This is understood in the framework of some aspect
of restricted quantum-classical correspondence \cite{Doron_cp1,doron_cp2,Doron_cp3,Doron_cp4}.
Let us denote the corrected eigenstates of $H$ by $\left|n'\right\rangle $.
To the first order in $u$,
\begin{equation}
\left|n\right\rangle '=\left|n\right\rangle +u\sum_{k\neq n}\left|k\right\rangle \cdot\frac{\left\langle k\left|S_{z}^{2}\right|n\right\rangle }{E_{n}-E_{k}}.
\end{equation}
The matrix element $\left\langle k\left|S_{z}^{2}\right|n\right\rangle $
can be calculated easily by using (\ref{eq:Sz_M}) up to the second
order in $\frac{1}{N}$. The result is
\begin{equation}
\left\langle k\left|S_{z}^{2}\right|n\right\rangle \approx\frac{1}{16}\left[d_{2}\left(n\right)\delta_{k,n+2}+d_{-2}\left(n\right)\delta_{k,n-2}+d_{0}\left(n\right)\delta_{k,n}\right]\label{eq:Sz2_matrix}
\end{equation}
where
\begin{eqnarray}
d_{2}\left(n\right) & = & 1-\frac{4\left(n+1\right)^{2}}{N^{2}}+\frac{2}{N}\nonumber \\
d_{-2}\left(n\right) & = & 1-\frac{4\left(n-1\right)^{2}}{N^{2}}+\frac{2}{N}\\
d_{0}\left(n\right) & = & 2+\frac{4}{N}-\frac{8n^{2}}{N^{2}}.\nonumber 
\end{eqnarray}
The energy difference is (\ref{eq:dE}) and therefore
\begin{equation}
\left|n\right\rangle '=\left|n\right\rangle +\frac{uN}{32}\left[d_{2}\left(n\right)\left|n+2\right\rangle -d_{-2}\left(n\right)\left|n-2\right\rangle \right].\label{eq:nn'}
\end{equation}
For small $n$ and $u$ relevant for the present work, it agrees with
the semiclassical result (\ref{eq:n'_clasi}). We would like to expand
the wavefunction in basis $\left|n\right\rangle '$. For this purpose,
we define the expansion coefficients
\begin{equation}
c'_{n}=\sum_{k}c_{k}\left\langle k|n\right\rangle .
\end{equation}
According to (\ref{eq:nn'}),
\begin{equation}
\begin{array}{ccl}
c'_{n} & \approx & c_{n}+\frac{uN}{32}\left[d_{2}\left(n\right)c_{n+2}-d_{-2}\left(n\right)c_{n-2}\right]\end{array}
\end{equation}
while the coefficients $c_{n}$ are given by (\ref{eq:cna}) ($\beta$
and $n_{max}$ are defined by (\ref{eq:betta}) and (\ref{eq:n_max}),
in Sec. V, $\beta=1$ and $n_{max}=0$, resulting in (\ref{eq:cn})).
Therefore,
\begin{equation}
c'_{n}=\left(\frac{2\beta}{\pi N}\right)^{\frac{1}{4}}e^{-\frac{1}{N}\beta\left(n-n_{max}\right)^{2}}\left[1+\frac{uN}{32}d_{2}\left(n\right)e^{-\frac{4}{N}\beta\left(n-n_{max}+1\right)}-\frac{uN}{32}d_{-2}\left(n\right)e^{+\frac{4}{N}\beta\left(n-n_{max}-1\right)}\right].
\end{equation}
In the leading order in $\frac{1}{N}$, $\frac{n}{N}\approx\frac{n_{max}}{N}$
and
\begin{equation}
c'_{n}=\left(\frac{2\beta}{\pi N}\right)^{\frac{1}{4}}e^{-\frac{1}{N}\beta\left(n-n_{max}\right)^{2}}\left\{ 1-\frac{u}{4}\left[\beta\left(1-\frac{4n_{max}^{2}}{N^{2}}\right)\left(n-n_{max}\right)+2\frac{n_{max}}{N}\right]\right\} ,
\end{equation}
and in the first order in $u$,
\begin{equation}
c'_{n}c'_{n+1}\approx\frac{\sqrt{2\beta}}{\sqrt{\pi N}}e^{-\frac{2\beta}{N}\left[\left(n-n_{max}\right)^{2}+\left(n-n{}_{max}\right)+\frac{1}{2}\right]}\left\{ 1-\frac{u}{2}\left[\beta\left(1-\frac{4n_{max}^{2}}{N^{2}}\right)\left(n-n_{max}+\frac{1}{2}\right)+2\frac{n_{max}}{N}\right]\right\} .
\end{equation}
The resulting correction to $\widetilde{S}_{m}$ is a of the form
\begin{equation}
\widetilde{S}{}_{m}^{\left(1\right)}=-\frac{\sqrt{2\beta}e^{-\frac{\beta}{N}}}{\sqrt{\pi N}}\cdot\frac{u}{2}\beta\left(1-\frac{4n_{max}^{2}}{N^{2}}\right)\int_{-\infty}^{\infty}xe^{-\left(Ax^{2}+Bx\right)}dx-\frac{u}{2}\left[2\frac{n_{max}}{N}+\frac{\beta}{2}\left(1-\frac{4n_{max}^{2}}{N^{2}}\right)\right]\widetilde{S}{}_{m}\label{eq:sm1}
\end{equation}
where $A,B$ are presented explicitly in App. B, Eq. (\ref{eq:AB})
and $x=n-n_{max}$. The integral can be solved by using 
\begin{eqnarray}
\int_{-\infty}^{\infty}xe^{-\left(Ax^{2}+Bx\right)}dx & = & \frac{-B}{2A}e^{\frac{B^{2}}{4A}}\sqrt{\frac{\pi}{A}}\label{eq:intx}
\end{eqnarray}
Therefore,
\begin{equation}
\widetilde{S}{}_{m}^{\left(1\right)}=-\frac{u}{2}\left[\beta\left(1-\frac{4n_{max}^{2}}{N^{2}}\right)\left(\frac{1}{2}-\frac{B}{2A}\right)+2\frac{n_{max}}{N}\right]\widetilde{S}{}_{m}.\label{eq:S1-1}
\end{equation}
Since $A$ and $B$ are of the same order of magnitude (see (\ref{eq:AB1})
and (\ref{eq:ABf})), this correction is typically small if $\frac{un_{max}}{N}$
is small. For the case $n_{max}=0$ discussed in Sec. V, $\widetilde{S}{}_{m}^{\left(1\right)}$
is negligible. However, in other cases (discussed in Sec. VI) it might
be important and then our approximation fails.

Now we calculate the second order correction for the case $\beta=1,\, n_{max}=0$
relevant for Sec. V.
\begin{eqnarray}
\left|n\right\rangle ' & = & \left|n\right\rangle +u\sum_{k\neq n}\left|k\right\rangle \cdot\frac{\left\langle k\left|S_{z}^{2}\right|n\right\rangle }{E_{n}-E_{k}}+u^{2}\sum_{l,k\neq n}\left|k\right\rangle \cdot\frac{\left\langle k\left|S_{z}^{2}\right|l\right\rangle \left\langle l\left|S_{z}^{2}\right|n\right\rangle }{\left(E_{n}-E_{k}\right)\left(E_{n}-E_{l}\right)}\nonumber \\
 &  & -u^{2}\sum_{k\neq n}\left|k\right\rangle \cdot\frac{\left\langle n\left|S_{z}^{2}\right|n\right\rangle \left\langle k\left|S_{z}^{2}\right|n\right\rangle }{\left(E_{n}-E_{k}\right)^{2}}-\frac{1}{2}u^{2}\left|n\right\rangle \cdot\sum_{k\neq n}\frac{\left\langle n\left|S_{z}^{2}\right|k\right\rangle \left\langle k\left|S_{z}^{2}\right|n\right\rangle }{\left(E_{n}-E_{k}\right)^{2}}.
\end{eqnarray}
According to (\ref{eq:Sz2_matrix}),
\begin{eqnarray}
\left|n\right\rangle ' & = & \left|n\right\rangle +\frac{uN}{32}\left[d_{2}\left(n\right)\left|n+2\right\rangle -d_{-2}\left(n\right)\left|n-2\right\rangle \right]\nonumber \\
 &  & +\frac{u^{2}N^{2}}{8\cdot16^{2}}\left[d_{2}\left(n\right)d_{2}\left(n+2\right)\left|n+4\right\rangle +d_{-2}\left(n\right)d_{-2}\left(n-2\right)\left|n-4\right\rangle \right]\nonumber \\
 &  & +\frac{u^{2}N^{2}}{4\cdot16^{2}}\left[d_{2}\left(n\right)d_{0}\left(n+2\right)\left|n+2\right\rangle +d_{-2}\left(n\right)d_{0}\left(n-2\right)\left|n-2\right\rangle \right]\nonumber \\
 &  & -\frac{u^{2}N^{2}}{4\cdot16^{2}}\left[d_{0}\left(n\right)d_{2}\left(n\right)\left|n+2\right\rangle +d_{0}\left(n\right)d_{-2}\left(n\right)\left|n-2\right\rangle \right]\nonumber \\
 &  & -\frac{u^{2}N^{2}}{8\cdot16^{2}}\left(d_{2}\left(n\right)d_{-2}\left(n+2\right)+d_{-2}\left(n\right)d_{2}\left(n-2\right)\right)\left|n\right\rangle \\
 & = & \left|n\right\rangle +\frac{uN}{32}\left[d_{2}\left(n\right)\left|n+2\right\rangle -d_{-2}\left(n\right)\left|n-2\right\rangle \right]\nonumber \\
 &  & +\frac{u^{2}N^{2}}{8\cdot16^{2}}\left[d_{2}\left(n\right)d_{2}\left(n+2\right)\left|n+4\right\rangle +d_{-2}\left(n\right)d_{-2}\left(n-2\right)\left|n-4\right\rangle \right]\nonumber \\
 &  & +\frac{u^{2}N^{2}}{4\cdot16^{2}}\left\{ \left[d_{2}\left(n\right)\left(d_{0}\left(n+2\right)-d_{0}\left(n\right)\right)\right]\left|n+2\right\rangle +\left[d_{-2}\left(n\right)\left(d_{0}\left(n-2\right)-d_{0}\left(n\right)\right)\right]\left|n-2\right\rangle \right\} \nonumber \\
 &  & -\frac{u^{2}N^{2}}{8\cdot16^{2}}\left(d_{2}\left(n\right)d_{-2}\left(n+2\right)+d_{-2}\left(n\right)d_{2}\left(n-2\right)\right)\left|n\right\rangle .\nonumber 
\end{eqnarray}
 namely,
\begin{eqnarray}
\left|n\right\rangle ' & \approx & \left|n\right\rangle +\frac{uN}{32}\left[\left(1-\frac{4\left(n+1\right)^{2}}{N^{2}}+\frac{2}{N}\right)\left|n+2\right\rangle -\left(1-\frac{4\left(n-1\right)^{2}}{N^{2}}+\frac{2}{N}\right)\left|n-2\right\rangle \right]\\
 &  & +\frac{u^{2}N^{2}}{8\cdot16^{2}}\left(1-\frac{4\left(n+1\right)^{2}}{N^{2}}-\frac{4\left(n+3\right)^{2}}{N^{2}}+\frac{4}{N}+\frac{4}{N^{2}}\right)\left|n+4\right\rangle \nonumber \\
 &  & +\frac{u^{2}N^{2}}{8\cdot16^{2}}\left(1-\frac{4\left(n-1\right)^{2}}{N^{2}}-\frac{4\left(n-3\right)^{2}}{N^{2}}+\frac{4}{N}+\frac{4}{N^{2}}\right)\left|n-4\right\rangle \\
 &  & -\frac{u^{2}N^{2}}{4\cdot16^{2}}\left\{ \frac{32\left(n+1\right)}{N^{2}}\left|n+2\right\rangle -\frac{32\left(n-1\right)}{N^{2}}\left|n-2\right\rangle \right\} \nonumber \\
 &  & -\frac{u^{2}N^{2}}{4\cdot16^{2}}\left(1-\frac{8\left(n^{2}+1\right)}{N^{2}}+\frac{4}{N}+\frac{4}{N^{2}}\right)\left|n\right\rangle .\nonumber 
\end{eqnarray}
Therefore,
\begin{eqnarray}
c_{n}' & \approx & \left(1-\frac{u^{2}N^{2}}{4\cdot16^{2}}\left(1-\frac{8\left(n^{2}+1\right)}{N^{2}}+\frac{4}{N}+\frac{4}{N^{2}}\right)\right)c_{n}\nonumber \\
 &  & +\frac{uN}{32}\left[\left(1-\frac{4\left(n+1\right)^{2}}{N^{2}}+\frac{2}{N}\right)c_{n+2}-\left(1-\frac{4\left(n-1\right)^{2}}{N^{2}}+\frac{2}{N}\right)c_{n-2}\right]\\
 &  & +\frac{u^{2}N^{2}}{8\cdot16^{2}}\left(1-\frac{4\left(n+1\right)^{2}}{N^{2}}-\frac{4\left(n+3\right)^{2}}{N^{2}}+\frac{4}{N}+\frac{4}{N^{2}}\right)c_{n+4}\nonumber \\
 &  & +\frac{u^{2}N^{2}}{8\cdot16^{2}}\left(1-\frac{4\left(n-1\right)^{2}}{N^{2}}-\frac{4\left(n-3\right)^{2}}{N^{2}}+\frac{4}{N}+\frac{4}{N^{2}}\right)c_{n-4}\\
 &  & -\frac{u^{2}N^{2}}{4\cdot16^{2}}\left[\frac{32\left(n+1\right)}{N^{2}}c_{n+2}-\frac{32\left(n-1\right)}{N^{2}}c_{n-2}\right]\\
 & = & \left(\frac{2}{\pi N}\right)^{\frac{1}{4}}e^{-\frac{n^{2}}{N}}\left\{ 1+\frac{uN}{32}\left[\left(1-\frac{4\left(n+1\right)^{2}}{N^{2}}+\frac{2}{N}\right)e^{-\frac{4n+4}{N}}-\left(1-\frac{4\left(n-1\right)^{2}}{N^{2}}+\frac{2}{N}\right)e^{\frac{4n-4}{N}}\right]\right.\nonumber \\
 &  & +\frac{u^{2}N^{2}}{8\cdot16^{2}}\left(1-\frac{4\left(n+1\right)^{2}}{N^{2}}-\frac{4\left(n+3\right)^{2}}{N^{2}}+\frac{4}{N}+\frac{4}{N^{2}}\right)e^{-\frac{8n+16}{N}}\nonumber \\
 &  & +\frac{u^{2}N^{2}}{8\cdot16^{2}}\left(1-\frac{4\left(n-1\right)^{2}}{N^{2}}-\frac{4\left(n-3\right)^{2}}{N^{2}}+\frac{4}{N}+\frac{4}{N^{2}}\right)e^{\frac{8n-16}{N}}\\
 &  & \left.-\frac{u^{2}N^{2}}{4\cdot16^{2}}\left[\frac{32\left(n+1\right)}{N^{2}}e^{-\frac{4n+4}{N}}-\frac{32\left(n-1\right)}{N^{2}}e^{\frac{4n-4}{N}}\right]-\frac{u^{2}N^{2}}{4\cdot16^{2}}\left(1-\frac{8\left(n^{2}+1\right)}{N^{2}}+\frac{4}{N}+\frac{4}{N^{2}}\right)\right\} .\nonumber 
\end{eqnarray}
Expending the exponent to the second order in $\frac{1}{N}$ yields
\begin{eqnarray}
c_{n}' & \approx & \left(\frac{2}{\pi N}\right)^{\frac{1}{4}}e^{-\frac{n^{2}}{N}}\left\{ 1-\frac{u^{2}N^{2}}{4\cdot16^{2}}\left(1-\frac{8\left(n^{2}+1\right)}{N^{2}}+\frac{4}{N}+\frac{4}{N^{2}}\right)-\frac{un}{4}\right.\nonumber \\
 &  & +\frac{u^{2}N^{2}}{4\cdot16^{2}}\left[1-\frac{12}{N}+\frac{24n^{2}+28}{N^{2}}\right]\nonumber \\
 &  & \left.-\frac{u^{2}N^{2}}{2\cdot16}\cdot\frac{2}{N^{2}}\right\} \\
 & = & \left(\frac{2}{\pi N}\right)^{\frac{1}{4}}e^{-\frac{n^{2}}{N}}\left\{ 1-\frac{un}{4}+\frac{u^{2}N}{4\cdot16}\left[-1+\frac{2n^{2}-2}{N}\right]\right\} \nonumber 
\end{eqnarray}
and
\begin{eqnarray}
c_{n}'c_{n+1}' & \approx & \frac{\sqrt{2}}{\sqrt{\pi N}}e^{-\frac{2n^{2}+2n+1}{N}}\left[1-\frac{u}{2}\left(n+\frac{1}{2}\right)\right.\\
 &  & \left.+\frac{u^{2}}{32}\left(-N+2n^{2}+2n+2n\left(n+1\right)\right)\right].\nonumber 
\end{eqnarray}
The second order correction to $\widetilde{S}_{m}$ is of the form
\begin{eqnarray}
\widetilde{S}_{m}^{\left(2\right)} & \approx- & \frac{u^{2}N}{32}\widetilde{S}_{m}+\frac{\sqrt{2}}{\sqrt{\pi N}}\cdot\frac{u^{2}}{8}\int\left(x^{2}e^{-\left(Ax^{2}+Bx\right)}+xe^{-\left(Ax^{2}+Bx\right)}\right)dx\label{eq:Smcor-1}
\end{eqnarray}
where $A$,$B$ are defined in (\ref{eq:AB}). The integral can be
calculated by using (\ref{eq:intx}) and
\begin{equation}
\int_{-\infty}^{\infty}x^{2}e^{-\left(Ax^{2}+Bx\right)}dx=\frac{1}{2A}e^{\frac{B^{2}}{4A}}\cdot\sqrt{\frac{\pi}{A}}.
\end{equation}
Therefore,
\begin{eqnarray}
\widetilde{S}_{m}^{\left(2\right)} & \approx & -\frac{u^{2}N}{32}\widetilde{S}_{m}+\widetilde{S}_{m}\cdot\frac{u^{2}}{16A}\cdot\left(1-\frac{B}{2}\right)\label{eq:Smcor-1-1}\\
 & \approx & -\frac{u^{2}N}{32}\widetilde{S}_{m}+\widetilde{S}_{m}\cdot\frac{u^{2}N}{32}.\left(1-\frac{1}{N}\right)\\
 & \approx- & \frac{u^{2}}{32}\widetilde{S}_{m}
\end{eqnarray}
and it is a small correction.

\section*{Appendix D}

In this appendix we calculate higher orders of the WKB expansion and
show that its contribution to the spectrum is not important. In the
WKB expansion \cite{Tabor}, one makes the ansatz
\begin{equation}
\psi\propto e^{iS\left(\varphi\right)/\hbar}
\end{equation}
where $S\left(\varphi\right)$ is the series
\begin{equation}
S\left(\varphi\right)=S_{0}\left(\varphi\right)+\hbar S_{1}\left(\varphi\right)+\hbar^{2}S_{2}\left(\varphi\right)+\ldots
\end{equation}
and
\begin{eqnarray}
\frac{\partial S_{0}}{\partial\varphi} & = & S_{x}\label{eq:S0}
\end{eqnarray}
\begin{equation}
S_{1}=\frac{1}{2}\ln S_{x}\label{eq:S1}
\end{equation}
\begin{equation}
\frac{\partial S_{2}}{\partial\varphi}=\frac{-1}{2\left(\frac{\partial S_{0}}{\partial\varphi}\right)}\left[\frac{\partial^{2}S_{1}}{\partial\varphi^{2}}+\left(\frac{\partial S_{1}}{\partial\varphi}\right)^{2}\right].\label{eq:S2}
\end{equation}
Here, $\hbar=\frac{1}{N}$. In Sec. III, we used the Bohr-Sommerfeld
quantization, namely, we demanded $S\left(\varphi\right)=S\left(\varphi+2\pi\right)+2\pi n\hbar$
in order to find the spectrum. In the present work, $\hbar=\frac{1}{N}$
is understood. There, we replaced $S$ by $S_{0}$ which is justified
only in the leading order in $\hbar.$ Finally, it turned out that
the spectrum contains terms of higher orders of $\hbar$ (\ref{eq:IH3})-(\ref{eq:E_BH2})
and therefore, the effects of $S_{1}$ and $S_{2}$ should be taken
into account as well. Fortunately, the contributions of $S_{1}$ and
$S_{2}$ are negligible as described in what follows. $S_{x}$ is
periodic in $\varphi$ so that $S_{1}$ does not contribute to the
spectrum.

In order to find the contribution of $S_{2}$, we first calculate
the derivatives of $S_{1}$:
\begin{equation}
\frac{\partial S_{1}}{\partial\varphi}=\frac{1}{2S_{x}}\frac{\partial S_{x}}{\partial\varphi}
\end{equation}
and
\begin{equation}
\frac{\partial^{2}S_{1}}{\partial\varphi^{2}}=\frac{-1}{2S_{x}^{2}}\left(\frac{\partial S_{x}}{\partial\varphi}\right)^{2}+\frac{1}{2S_{x}}\frac{\partial^{2}S_{x}}{\partial\varphi^{2}}.
\end{equation}
Hence,
\begin{equation}
\frac{\partial S_{2}}{\partial\varphi}=\frac{-1}{4S_{x}^{2}}\left[\frac{-1}{2S_{x}}\left(\frac{\partial S_{x}}{\partial\varphi}\right)^{2}+\frac{\partial^{2}S_{x}}{\partial\varphi^{2}}\right].\label{eq:dS2Sx}
\end{equation}
In what follows, all calculations are performed to the order $\hbar^{2}$.
We substitute $S_{x}$ of (\ref{eq:Sx_sec}) and find
\begin{equation}
\frac{\partial S_{x}}{\partial\varphi}=\left(\frac{1}{4}-H^{2}\right)\left[u\sin\left(2\varphi\right)+8u^{2}H\sin^{3}\varphi\cos\varphi\right],
\end{equation}
and
\begin{equation}
\frac{\partial^{2}S_{x}}{\partial\varphi^{2}}=\left(\frac{1}{4}-H^{2}\right)\left[2u\cos\left(2\varphi\right)+8u^{2}H\left(3\sin^{2}\varphi\cos^{2}\varphi-\sin^{4}\varphi\right)\right].
\end{equation}
Therefore, to the second order in $u$,
\begin{equation}
\frac{\partial S_{2}}{\partial\varphi}=\frac{-1}{4S_{x}^{2}}\left(\frac{1}{4}-H^{2}\right)\left[\frac{-u^{2}}{2S_{x}}\left(\frac{1}{4}-H^{2}\right)\sin^{2}\left(2\varphi\right)+2u\cos\left(2\varphi\right)+8u^{2}H\left(3\sin^{2}\varphi\cos^{2}\varphi-\sin^{4}\varphi\right)\right].
\end{equation}
Assuming $u\ll H$ we find $\frac{1}{S_{x}^{2}}\approx\frac{1}{H^{2}}\left(1+\frac{2}{H}u\left(\frac{1}{4}-H^{2}\right)\sin^{2}\varphi\right)$,
leading to
\begin{eqnarray}
\frac{\partial S_{2}}{\partial\varphi} & = & \frac{-1}{4H^{2}}\left(\frac{1}{4}-H^{2}\right)\left[\frac{u^{2}}{2H}\left(\frac{1}{4}-H^{2}\right)\left(\sin^{2}\left(2\varphi\right)+8\cos\left(2\varphi\right)\sin^{2}\varphi\right)\right.\\
 &  & \left.+2u\cos\left(2\varphi\right)+8u^{2}H\left(3\sin^{2}\varphi\cos^{2}\varphi-\sin^{4}\varphi\right)\right].
\end{eqnarray}
Therefore,
\begin{equation}
\bar{\delta}=\frac{\hbar^{2}}{2\pi}\left(S_{2}\left(\varphi+2\pi\right)-S_{2}\left(\varphi\right)\right)=\frac{-3u^{2}}{16H^{3}N^{2}}\left(\frac{1}{4}-H^{2}\right)^{2}.
\end{equation}
This should be added to the right hand side of (\ref{eq:IH3}), resulting
in a contribution of 
\begin{equation}
\delta'=\frac{3u^{2}N}{16n^{3}}\left(\frac{1}{4}-\frac{n^{2}}{N^{2}}\right)^{2}
\end{equation}
to the spectrum (\ref{eq:E_BH2-1}). The approximation leading to
this term is not valid for small $n$ (see (\ref{eq:En}) where $H$
is of order $u$). To find an estimate for the correction in this
regime we repeat the calculation for $I=0$. If $n=0$,
\begin{equation}
\begin{array}{ccl}
S_{x} & \approx & -\frac{u}{8}+\frac{1}{4}u\sin^{2}\varphi\end{array}\label{eq:Sx_sec-1}
\end{equation}
and the derivatives are  $\frac{\partial S_{x}}{\partial\varphi}=\frac{1}{4}u\sin\left(2\varphi\right),$
$\frac{\partial^{2}S_{x}}{\partial\varphi^{2}}=\frac{1}{2}u\cos\left(2\varphi\right)$.
Therefore, to the second order in $u$,
\begin{eqnarray}
\frac{\partial S_{2}}{\partial\varphi} & = & \frac{-1}{4u\left(-\frac{1}{8}+\frac{1}{4}\sin^{2}\varphi\right)^{2}}\left[\frac{-1}{2\left(-\frac{1}{8}+\frac{1}{4}\sin^{2}\varphi\right)}\left(\frac{1}{4}\sin\left(2\varphi\right)\right)^{2}+\frac{1}{2}\cos\left(2\varphi\right)\right]\\
 & = & \frac{4}{u\cos^{3}\left(2\varphi\right)}\left[\sin^{2}\left(2\varphi\right)-2\cos^{2}\left(2\varphi\right)\right].\label{eq:ds2}
\end{eqnarray}
This expression is antisymmetric with respect to $2\varphi=\frac{\pi}{2}+\alpha\rightarrow2\varphi=\frac{\pi}{2}-\alpha$.
Therefore the integral for $S_{2}$ vanishes. The above estimates
are only for part of the spectrum. Therefore, we turn to a numerical
estimate.

In Fig. 5, we present the numerically calculated deviations in the
spectrum originating of $S_{2}$ and show that it is small for the
parameters of Figs. 2-3. The calculation of the spectrum presented
in Fig. 5 was carried out by iterations as described in what follows:
\begin{enumerate}
\item For each $n$, $S_{x}\left(\varphi\right)$ was calculated according
to (\ref{eq:Sx}) where $H$ is replaced by the spectrum $E_{n}^{\left(2\right)}$
of (\ref{eq:E_BH2-1}). 
\item $S_{2}\left(\varphi\right)$ was found by substitution of $S_{x}$
in (\ref{eq:dS2Sx}) and integration over $\varphi$.
\item The term $\frac{\hbar^{2}}{2\pi}\left(S_{2}\left(\varphi+2\pi\right)-S_{2}\left(\varphi\right)\right)$
was added to the RHS of (\ref{eq:IH3}), which we solved numerically
to obtain a corrected spectrum $\widetilde{E}{}_{n}^{\left(2\right)}$.
\item We repeated steps 1-3 where $H$ in $S_{x}$ is replaced by $\widetilde{E}{}_{n}^{\left(2\right)}$
until conversion.
\item We Multiplied the resulting spectrum by $2JN$ and added the constant
$C_{N}$ to be able to compare with the exact BH spectrum.
\end{enumerate}
\selectlanguage{american}%
\begin{figure}[H]
\selectlanguage{english}%
(a)\qquad{}\qquad{}\qquad{}\qquad{}\qquad{}\qquad{}\qquad{}\qquad{}\qquad{}\qquad{}\qquad{}\qquad{}(b)

\includegraphics[scale=0.45]{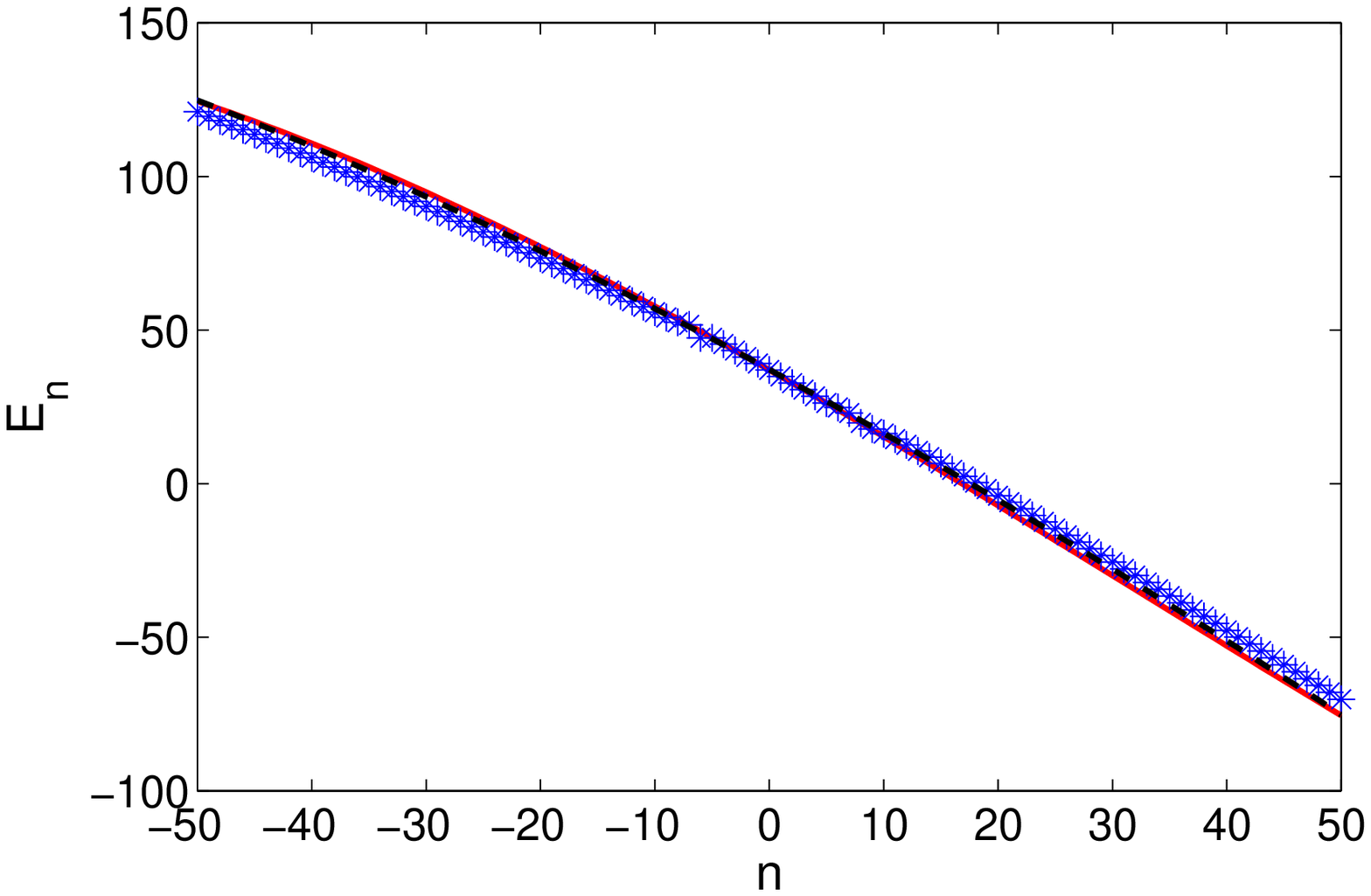}\includegraphics[scale=0.45]{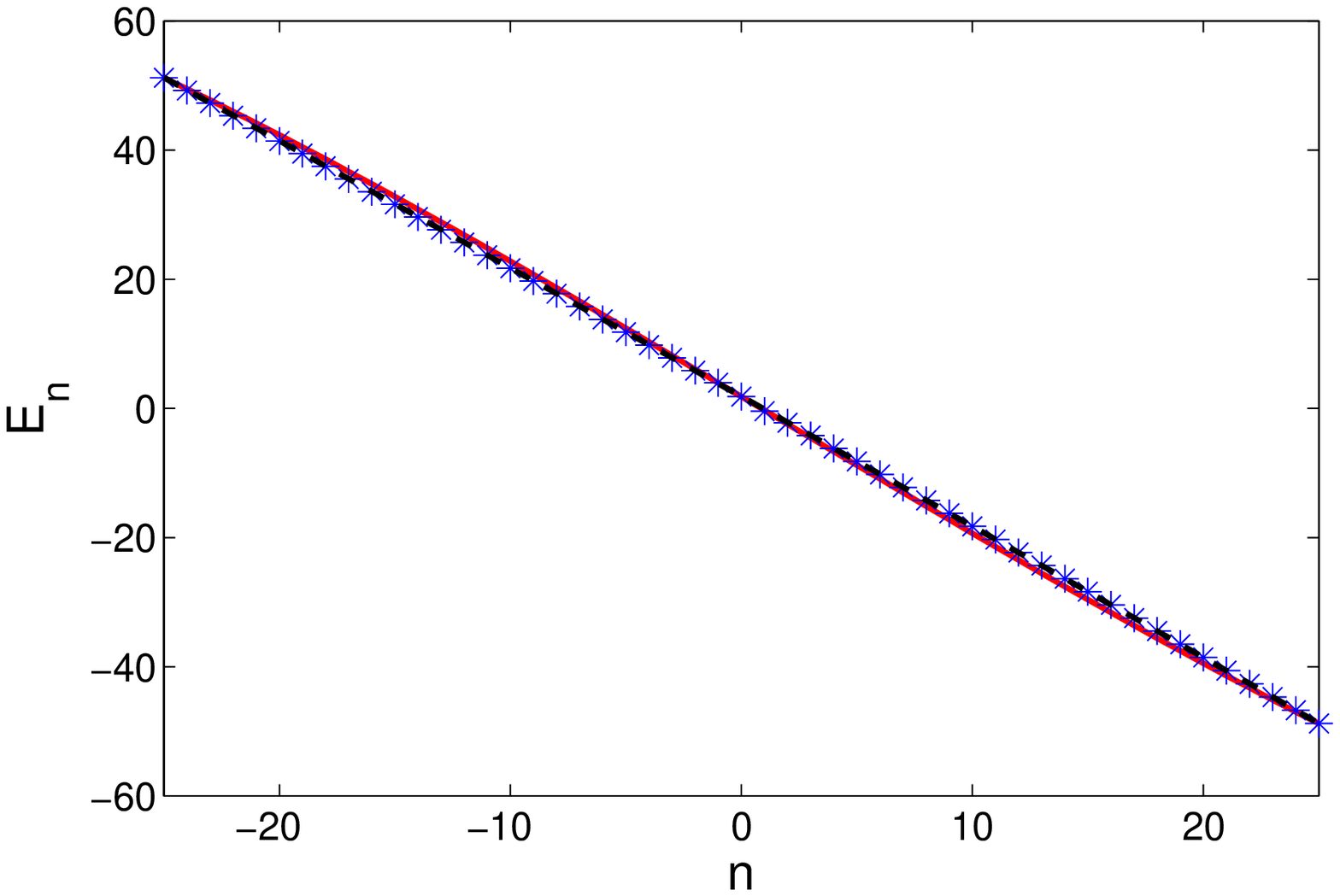}

\selectlanguage{american}%
\caption{\selectlanguage{english}%
(Color online) Spectrum of the BH Hamiltonian, The red lines represent
the spectrum (\ref{eq:E_BH2}) that was used in the calculation of
the dynamics and the blue stars represent the spectrum which was obtained
numerically by taking into account contributions up to order $\hbar^{2}$
in the semiclassical approximation, as described in the text. The
exact spectrum of the BH Hamiltonian (obtained by diagonalization
\foreignlanguage{american}{(\ref{eq:H_BH})) appears in black dashed
line.} (a) $J=1$, $N=100$ and $u=\frac{1}{2}$. (b) $J=1$, $N=50$
and $u=\frac{1}{20}$.\selectlanguage{american}
}
\end{figure}

\selectlanguage{english}%

\section*{Appendix E}

In this appendix, we calculate the eigenstates in the semiclassical
approximation 
\begin{equation}
\left|n'\right\rangle =\frac{1}{\sqrt{2\pi}}e^{iS_{0}\left(\varphi\right)/\hbar}\label{eq:semi_wave}
\end{equation}
 and show that it can be approximated by the eigenstates of $S_{x}$
as was done in Sec. V and VI. According to (\ref{eq:S0}) and (\ref{eq:Sx_first}),
in the first order in $u$,
\begin{equation}
S_{0}=\frac{k}{N}\varphi-\frac{u}{4}\left[\frac{1}{4}-\frac{k{}^{2}}{N^{2}}\right]\sin\left(2\varphi\right).\label{eq:s0f}
\end{equation}
The eigenstates of $S_{x}$ (obtained by substituting (\ref{eq:s0f})
with $u=0$ in (\ref{eq:semi_wave})) are $e^{in\varphi}$. These
are denoted by $\left|n\right\rangle $ of (\ref{eq:S_x_eigen}).
The overlap between $\left|k\right\rangle '$ and $\left|n\right\rangle $
is
\begin{equation}
\left\langle n|k\right\rangle '=\frac{1}{2\pi}\int_{0}^{2\pi}e^{i\left[\varphi\left(k-n\right)-\widetilde{C}_{2}\sin\left(2\varphi\right)\right]}d\varphi
\end{equation}
where $\widetilde{C}_{2}=\frac{uN}{4}\left[\frac{1}{4}-\frac{k^{2}}{N^{2}}\right]$
. In order to solve the integral, we expand to series of Bessel functions:
\begin{equation}
e^{-i\widetilde{C}_{2}\sin\left(2\varphi\right)}=\sum_{l=0}^{\infty}J_{l}\left(\widetilde{C}_{2}\right)e^{-2il\varphi}+\sum_{l=1}^{\infty}\left(-1\right)^{l}J_{l}\left(\widetilde{C}_{2}\right)e^{2il\varphi}
\end{equation}
and obtain
\begin{eqnarray}
\left\langle n|\left(n+2l\right)\right\rangle ' & = & J_{l}\left(\widetilde{C}_{2}\right)\\
\left\langle n|\left(n-2l\right)\right\rangle ' & = & \left(-1\right)^{l}J_{l}\left(\widetilde{C}_{2}\right)
\end{eqnarray}
for positive integer $l$. Since $\widetilde{C}_{2}$ is small, the
Bessel functions can be approximated by $J_{l}\left(\widetilde{C}_{2}\right)\sim\frac{1}{l!}\left(\frac{\widetilde{C}_{2}}{2}\right)^{l}$,
so that the overlap is substantial only for small values of $l$ and
\begin{equation}
\left|n\right\rangle '\approx\left|n\right\rangle +\frac{1}{2}\widetilde{C}_{2}\left[\left|n+2\right\rangle -\left|n-2\right\rangle \right].\label{eq:n'_clasi}
\end{equation}
This result reduces to (\ref{eq:nn'}) for small $n$ and contribute
only small corrections to the dynamics, as was shown in App. C.

\selectlanguage{american}%
\bibliographystyle{h-physrev}
\addcontentsline{toc}{section}{\refname}\bibliography{C:/Users/Hagar/Dropbox/harmonic_paper/referances}
\selectlanguage{english}

\end{document}